\documentclass[12pt]{JHEP3}

\def\f{\frac}

\def\l{\lambda}

\def\m{\mu}

\def\nn{\nonumber}
\def\o{\omega}

\def\T{\tau}

\newcommand{\be}{\begin{equation}}
\newcommand{\ee}{\end{equation}}
\newcommand{\bea}{\begin{eqnarray}}
\newcommand{\eea}{\end{eqnarray}}
\newcommand{\barr}{\begin{array}}
\newcommand{\earr}{\end{array}}

\title{Sum rules and three point functions}
\author{ Justin R. David  and  Somyadip
Thakur   \\
  Centre for High Energy Physics,
Indian Institute of Science,\\ C.V. Raman Avenue, Bangalore 560012, India. \\
\email{justin, somyadip@cts.iisc.ernet.in}\\

}

\abstract{
Sum rules constraining the R-current spectral densities  are derived 
holographically for the case of D3-branes, M2-branes and M5-branes all at 
finite chemical potentials. 
In each of the cases the sum rule relates a certain integral of the spectral density 
over the frequency  to terms  
which depend both on long distance physics, hydrodynamics and 
short distance physics of the theory. 
The terms which which depend on the short distance physics
 result from  the presence of certain chiral primaries  in the 
OPE of two R-currents which are turned on at finite chemical potential. 
Since these sum rules contain information of the OPE they 
provide an alternate method to obtain  the structure constants of the 
two R-currents and the chiral primary.  As a consistency check we show that the 3 point function 
derived from the sum rule precisely matches with that obtained using 
Witten diagrams. }

\begin{document}

\section{Introduction}

Sum rules  constraining spectral densities 
in  any quantum field theory provide useful information of the theory. 
For example  the shear  and bulk sum rules in QCD is a consequence of 
the hydrodynamic behaviour at large distances and asymptotic 
freedom at short distances 
\cite{Kharzeev:2007wb,Karsch:2007jc,Romatschke:2009ng,Meyer:2010ii,Meyer:2010gu}. 
Similarly the Ferrell-Glover sum rule in BCS  theory plays an important 
role in determining the  skin depth
of superconductors \cite{Ferrell:1958zza,PhysRevLett.2.331}. 

Recently there has been interest in
modeling systems that are studied in condensed matter 
physics  as well as QCD  using  the gauge gravity duality. 
Clearly sum rules derived using the  AdS/CFT framework for these systems can
provide important constraints on these models. On a more conceptual level
since sum rules are a consequence of unitarity and causality of the field theory, 
they   will be useful to study how these properties are encoded 
in the bulk.
The study of sum rules in AdS/CFT was initiated in \cite{Romatschke:2009ng}
in which the shear spectral sum rule was derived and  verified numerically for the 
${\cal N}=4$ theory at strong coupling. 
The shear  and bulk sum rules in non-conformal theories  dual to 
the Chambilin-Real backgrounds were obtained  in 
\cite{Springer:2010mf,Springer:2010mw}. The R-charge sum rule for ${\cal N}=4$
Yang-Mills was derived holographically in \cite{Baier:2009zy}.

Using the gauge/gravity framework,  the derivation of the sum rules for the 
strongly coupled dual can be cast into a problem of establishing   the 
analytic behaviour of  solutions of the differential equations which determine 
the relevant retarded Green's function \cite{Gulotta:2010cu}. 
By studying  properties of these differential equations,  a proof for various
sum rules including the shear sum rule for ${\cal N}=4$ Yang-Mills was 
provided in \cite{Gulotta:2010cu}. 
In \cite{David:2011hy}, this proof for the shear sum rules 
was developed  and extended to  systems with 
chemical potential.  It was noticed there that  the shear  sum rules 
are modified due to the short distance properties of the theory. 
It was shown that  appropriate  scalar operators   in the 
stress tensor operator product expansion (OPE)
 gain expectation values in the presence of the chemical potential. 
This results in the modified shear sum rule for the case of D3, M2 and M5-branes
when compared to the situation in the absence of chemical potentials.  

In the present paper we derive the R-current
spectral sum rules at finite chemical potential for the dual of the D3, M2 and M5-brane
theory. 
The differential equations which determine the R-current correlator 
are a set of coupled equations unlike the situation for the shear correlator which was 
determined by a single equation.  
We show that  the R-charge  sum rules contain both a term determined by hydrodynamics
at long distance: the conductivity,  as well as a term whose origin is 
from short distance physics: the OPE..  We then examine the 
term due to the presence of scalars in the OPE and obtain the structure  constants of 
two R-currents and the scalar.  We show that the structure constants evaluated 
from the sum rule agrees precisely with that evaluated using
Witten diagrams. 

We now briefly  describe the structure of the R-current spectral sum rules derived in this paper. 
The formulae are of the form
\begin{equation}
\label{ucsumi}
\int_{-\infty}^\infty \frac{d\omega}{\pi\omega}
\left( \rho^i_T(\omega) - \rho^i_{T=0} (\omega) \right) 
=  \lim_{\omega\rightarrow 0} \omega {\rm Im} \, \sigma^i(\omega) 
-  C_{ii}^{\;\;\hat k} \langle {\cal O}^k \rangle_{T}, 
\end{equation}
where  $\rho^{i}(\omega) = {\rm Im}\, G^i$ and   $G^i$ is the 
retarded Green's function of the $x$ component of the R-current $J_x^i$.  $\sigma^i$ refers to  the 
corresponding conductivity.  $C_{ii}^{\;\;\hat k}$ is the structure constant of two R-currents 
and the scalar ${\cal O}^{\hat k}$. The index  $\hat k$  is summed
over all the operators of appropriate dimensions in the theory.  The expectation value of this scalar is 
taken in the thermal state at temperature $T$. 
Notice that the first term in the R-current sum 
rule is determined by hydrodynamics, while the second term is determined by 
the short distance physics of the theory. 
For the case of D3-branes the operators ${\cal O}^{\hat k}$ are chiral primaries of dimension $2$.
While for the case of M2, M5-branes they are chiral primaries  
 of dimension $1$ and $4$ respectively. 
It is clear from (\ref{ucsumi})
that once the one knows the expectation values of these primaries, the structure
constants can be extracted. In each of the situations we verify that the structure constants
obtained from the sum rule agrees with that evaluated from Witten diagrams.

The organization of the paper is as follows:
In the next section we first recall
the general analytical properties of the 
retarded Green's function  required to derive sum rules for the corresponding spectral 
densities. We then go over the details of the derivation of the R-current spectral sum rules for the 
dual of ${\cal N}=4$ Yang-Mills at finite chemical potential. 
In section 3, we examine the origin of the high frequency 
contribution to the sum rules. We show that these terms occur due to presence 
of chiral primary operators of dimension 2 in the OPE of two R-currents. 
From this we derive the structure constants of two R-currents with these chiral primaries.
As a consistency check,  
these constants are then compared to that obtained from Witten diagrams and shown
to agree.  In section 4, the analysis is repeated for the case of M2 and M5-branes. 
The discussion in this section is brief  and  only the results are highlighted.
Section 6 contains our conclusions. 
Appendix A discussed the evaluation  of the 3 point and 2 point functions 
using Witten diagrams which are necessary to compare with that 
obtained using the sum-rules. 
Appendix B provides the necessary details of the charged M2 and M5-brane solution.

\section{R-charge sum rules for the  D3-brane}

Before we begin the derivation of the spectral sum rules for the D3-branes we briefly 
state the analytic properties  of the Green's function in the complex $\omega$-plane
which are required to derive sum rules. 
Consider a function $ G(\omega)$ which satisfies the following two properties 
in the complex $\omega$-plane. 
\begin{enumerate}
 \item  $G(\omega)$ is holomorphic in the upper half $\omega$-plane, including the 
real axis. \\
\item $\lim_{|\omega|\rightarrow \infty}$ = 0 for ${\rm Im}\, \omega >0$.
\end{enumerate}
We refer to these properties as `property 1'and property 2' respectively. 
Then one can show using Cauchy's theorem that 
\begin{equation}
 G(0) = \lim_{\epsilon\rightarrow 0^+} \int_{-\infty}^\infty \frac{d\omega}{\pi} \frac{\rho(\omega)}{\omega - i\epsilon},  
\end{equation}
where $\rho(\omega) = {\rm Im }\, G(\omega)$. 
The details of this analysis can be found in \cite{David:2011hy}. 
In this section we examine the retarded Green's function corresponding to the 
R-current correlators for ${\cal N}=4$ Yang-Mills
at strong coupling and construct a regulated Green's function which satisfies 
both the above properties and thus derive the spectral sum rule. 

We first  introduce the gravity dual of ${\cal N}=4$ super Yang-Mills 
at finite temperature and finite chemical potential.  We examine the situation 
in which the chemical potentials corresponding to the three Cartans of the 
$SO(6)$ R-symmetry is turned on.  The gravity dual of this system is given by 
the R-charged black hole of Behrndt, Cvetic and Sabra \cite{Behrndt:1998jd}. 
We study the retarded Green's function of the R-symmetry currents 
in this background. We will focus on the following diagonal correlator:
\begin{equation}
 G^{i}(t, \vec x) = i \theta( t) \langle [J^i_x (t, \vec x), J_x^i( t, \vec x) ] \rangle,   
\end{equation}
 where $J^i_x$ is the $x$ component of the $i$-th R-symmetry current and 
$i \in \{1, 2, 3 \}$. 
This is done by first  obtaining the coupled set of differential equations 
for the fluctuations of the gauge fields dual to the R-symmetry currents
in the R-charged black hole. 
We then  evaluate  the retarded Green's function using the 
standard prescription developed in \cite{Policastro:2002se,Kovtun:2005ev}. 
Examining these set of differential equations we will show that the
Green's function satisfies the two properties necessary for the derivation 
of the sum rule. 
We will show that the RHS of sum rule  
depends on two terms as given in ( \ref{ucsumi}).  
The first term depends on the zero frequency behaviour of the Green's function and 
can be understood in terms of the hydrodynamic properties of the field theory. 
The second term arises due to the high frequency behaviour of the Green's function. 
In the next section it will be shown   that the high frequency terms in the sum rule result from 
scalars corresponding to chiral primary operators gaining  expectation 
values at finite chemical potential.

\subsection{Green's function from gravity}

The R-charged D3-brane solution of \cite{Behrndt:1998jd} in 5 dimensions is given by 
\begin{eqnarray}
\label{chgmet}
 ds^2_5 &=& - { \cal H}^{-2/3}{(\pi T_0 L)^2 \over u}\,f \, dt^2 
+  { \cal H}^{1/3}{(\pi T_0 L)^2 \over u}\, \left( dx^2 + dy^2 + dz^2\right)
+ {\cal H}^{1/3}{L^2 \over 4 f u^2} du^2, \, \nonumber  \\ \nonumber
f(u) &=& { \cal H} (u) - u^2  \prod_{i=1}^3 (1+k_i)\,, 
\;\;\;\;\; H_i = 1 + k_i u \,, \;\;\;\;\; 
k_i  \equiv {q_i\over r_H^2}\,, \;\;\;\;\; T_0 =\frac{ r_+}{\pi L^2},
\label{identif} \\ 
u &=& \frac{r_+^2}{r^2}, \quad  {\cal H} = ( 1+ k_1u)( 1+k_2u) (1+k_3u) .
\end{eqnarray}
The solution also involves scalars and vectors which have the following 
background values. 
\begin{equation}
\label{scalval}
 X^i = \frac{ {\cal H}^{1/3} }{ H_i ( u) }, \quad 
A_t^i = \frac{\tilde  k_i u}{  H_i(u) }, \quad
\tilde k_i = \frac{ \sqrt{q_i}}{L} \prod_{i=1}^3 ( 1 + k_i)^{1/2} .
\end{equation}
This background solves the equation of motion of the action which given by 
\begin{eqnarray}
\label{cgaction}
 S &=& \frac{N^2}{8\pi^2 L^3} \int d^5 x\sqrt{-g} {\cal  L}, \\ \nonumber
\mathcal{L} &=&  R + \frac{2}{L^2} V - 
{\frac{1}{2 }} \tilde G_{ij} F_{\mu\nu}
^i F^{\mu\nu j}- { \tilde G}_{ij} \partial_{\mu} 
X^i  \partial^\mu
X^j +{\frac{1}{24\sqrt{-g}} \epsilon^{\mu\nu\rho\sigma \lambda} {\cal{C}}_{ijk}
F_{\mu\nu}^i F_{\rho\sigma}^j A_{\lambda}^k}, \nonumber\\ 
\end{eqnarray}
where $F^{i}_{\mu\nu}, i =1, 2, 3$ are the field-strengths for the 
three Abelian gauge fields corresponding to the R-currents of the boundary theory. 
The three scalar fields $X^i$'s are subject to the 
constraint $X^1X^2X^3 =1$ and ${\cal{C}}_{ijk}$ are coefficients 
totally symmetric  in the R-symmetry indices. 
The metric on the scalar manifold is given by 
\begin{equation}
 \tilde G_{ij} = \frac{1}{2} {\rm diag} \left\{ (X^{1})^{-2},  (X^{2})^{-2},  (X^{3})^{-2} \right\}.
\end{equation}
The scalar potential is given by 
\begin{equation}
 {\cal V} = 2 \left( \frac{1}{X^1} +   \frac{1}{X^2} + \frac{1}{X^3} \right) .
\end{equation}
The thermodynamic  properties of this black hole have been obtained in \cite{Son:2006em}. 
Note that unlike in \cite{Son:2006em}, 
we are working with units such that the gauge fields $A^i$ are dimensionless. 
The Hawking temperature $T_H$, entropy density $s$ , energy density $\epsilon$, pressure $P$, 
charge densities $\rho_i$  and the conjugate chemical potentials $\mu_i$  are given by 
\begin{eqnarray}
\label{cthermv}
 T_H = \frac{ 2 +k_1 + k_2 + k_3 - k_1k_2k_3}
{ 2 \sqrt{( 1+ k_1) ( 1+ k_2) ( 1+ k_3) } } T_0, &\qquad& 
s =  \frac{\pi^2 N^2 T_0^3}{2} \prod_{i=1}^3 ( 1 + k_i)^{1/2}, 
\\ \nonumber
\epsilon = \frac{ 3\pi^2 N^2 T_0^4}{8} \prod_{i=1}^3 ( 1 + k_i) ,  &\qquad& 
P = \frac{\pi^2 N^2 T_0^4}{ 8}  \prod_{i=1}^3 ( 1 + k_i), \\ \nonumber
\rho_i = \frac{\pi N^2T_0^3}{4 L} \sqrt{k_i}  \prod_{i=1}^3 ( 1 + k_i),  &\qquad& 
\mu_i = \frac{\pi T_0 L \sqrt{k_i}}{( 1+ k_i)}  \prod_{i=1}^3 ( 1 + k_i)^{1/2}. 
\end{eqnarray}
The thermodynamical stability condition of this black hole is given by 
\begin{equation}
 2 -k_1-k_2 - k_3 + k_1 k_2k_3>0.
\end{equation}
It is also useful to write down the explicit equations of motion that follow from the action in 
(\ref{cgaction}).  The Einstein equations are 
\begin{eqnarray}\label{einom}
 R_{\mu\nu}- \left(F^{2}_{\mu \nu} - {1 \over 6} g_{\mu\nu} F^2 \right)
   - \partial_{\mu} X^i \partial_{\nu} X^j \, \tilde G_{ij} 
   +{4 \over 3 L^2}  g_{\mu\nu}(\frac{1}{X^1}+\frac{1}{X^2}+\frac{1}{X^3}) = 0.
\end{eqnarray}
with $F^2_{\mu\nu} \equiv \tilde G_{ij} F_{\mu\rho}^{i} F^{j}_{\nu\lambda} \,
g^{\rho\lambda}$. 
The equations of motion for the gauge fields are given by 
\begin{eqnarray}\label{gom}
 {1 \over \sqrt{g}} \, \partial_{\mu} 
\left( \sqrt{g} \, \tilde G_{ij} F^{j\, \mu\nu} \right)
+3\kappa \epsilon^{\nu \alpha \beta \gamma \lambda} F_{\alpha \beta}^j
F_{\gamma \lambda}^k {\cal{C}}_{ijk} =0, 
\end{eqnarray}
where $\kappa = \frac{1}{48}$. 
Finally the equations of motion for the scalars are given by
\begin{eqnarray} \label{som}
{1 \over \sqrt g} \partial_{\mu} ( \sqrt g g^{\mu\nu} {{ G}}_{ij}
\partial_{\nu} X^j ) -{1 \over 4} \partial_i \tilde G_{jk}
F^{j}_{\mu\nu} F^{k\, \mu\nu} + \frac{1}{L^2} \partial_i V(X) =0\ ,
\end{eqnarray}
Here $\partial_i$ refers to derivative with respective to the scalar $X^i$. 

To study  the retarded Green's function for the currents $J_x^i$ we need to 
obtain the equations of motion of the fluctuations for the dual  gauge field $A_x^i$. Further more
we are interested in the Green's function at zero momentum, therefore we can 
restrict our attention to fluctuations which have only time dependence. 
It  can be shown that it is consistent to turn on the following fluctuations
\begin{eqnarray}
\label{fluct}
A_x^i = A_x ^{i(0)} +  a_x^i( u, t), \qquad A_l^i = A_l^{i(0)},  \\ \nonumber
g_{xt} = g_{xt}^{(0)} + h_{xt}( u, t) , \qquad g_{lt} = g_{lt}^{(0)}, \\ \nonumber
g_{lm} = g_{lm}^{(0)}, \qquad X^i = X^{i(0)}. 
\end{eqnarray}
Here $l \in\{y, z, t, u \}$ and the superscript ${}^{(0)}$ refers to the background values 
given in (\ref{chgmet}) and (\ref{scalval}).  Note that $u$ refers to the radial direction. 
The consistency of setting  several of the fields to their background values 
in (\ref{fluct})  
can be shown by examining the equations of motion 
given in (\ref{einom}), (\ref{gom}) and (\ref{som}),
We now define 
\begin{equation}
T(u) e^{i\omega t} , \qquad   a_x^i (u, t)  = \phi^i (u) e^{i\omega t } . 
\end{equation}
Here we have used the time translational invariance of the problem to study fluctuations 
of a given frequency $\omega$. Note that the same analysis can be carried out for any of the 
spatial directions. By the symmetry of the problem it is easy to see that the results will 
be independent of any of the direction. 
Then examining the $xu$ component of the Einstein equation given in (\ref{einom})
to first order in the fluctuations  we obtain 
the following constraint
 \begin{eqnarray}
\label{constraint}
T' +\sum_{i=1}^3\left(  \frac{ u}{{ \cal H}}(1+k_{i}) \frac{m_{i}}{\pi T_0 L}  \phi^{i} \right)  =0,
\end{eqnarray}
where the prime refers to derivative with respect to $u$. 
We now examine the $x$ component of the equations of motion of the  three gauge fields given
in (\ref{gom}) to obtain 
\begin{eqnarray}
\label{gageqn}
 \mu_i (1+k_i) T' + 
\left(  \frac{f H_{i}^{2}}{\cal H} \frac{d {\phi}^{i  }}{du} \right)^{\prime} 
+ \frac{\tilde{\omega}^2}{uf}  H_{i}^{2} {\phi }^{i}= 0, 
\end{eqnarray}
where we have defined  $\tilde{\omega}=\frac{\omega}{2 \pi T_0}$.
Note that in the equation in (\ref{gageqn}) there is no summation over the index $i$. 
Substituting the $T'$ from (\ref{constraint}) in the equation (\ref{gageqn}) we obtain the 
closed set of equations for the gauge field fluctuations. 
These are given by 
\begin{eqnarray}
\label{jjeqn}
 \frac{d }{du} \left(  \frac{f H_{i}^{2}}{\cal H} \frac{d { \phi}^{i}}{du}  \right) 
  -(1+k_i)m_i \sum_{j=1}^3\left(  \frac{ u}{{\cal H}}(1+k_{j}) m_{j} { \phi}^{j}\right)   
+ \frac{\tilde{\omega}^2}{uf}  H_{i}^{2} { \phi}^{i}=0. 
\end{eqnarray}
It is useful to rewrite the above equation in terms of the conventional radial coordinate $r$
which is  related to  $u$ by 
\begin{equation}
 u = \frac{r_+^2}{r^2}. 
\end{equation}
Then the equation (\ref{jjeqn}) is given by 
\begin{eqnarray}
\label{gageqnr}
{\phi}^{i\prime\prime}+
\left( \ln(\frac{FH_i^2}{{\cal H} })'+\f{1}{r} \right){\phi}^{i\prime}+\frac{\omega^2{\cal H} }{F^2}{ \phi}^{i}-
(1+k_i) \frac{m_i}{H_i^2} \sum_{j=1}^3
\left( \frac{4 r_+^6}{ r^6 L^2 F}(1+k_{j}) m_{j} {\phi}^j \right) =0,  \nonumber \\
\end{eqnarray}
where the prime now denotes derivative with respect to $r$ and 
\begin{equation}
\label{deff}
F = \frac{r^2}{L^2} f . 
\end{equation}
We will also need the fact that the equations of motion for the 
gauge field fluctuations given in (\ref{gageqnr}) can be obtained from 
the following action
\begin{eqnarray}
\label{varact2}
 S_{\phi}= \int_{r_h}^{\infty}dr \frac{F r H_i^2}{{\cal H}}
\left(  \frac{d \phi^{i *}}{dr} \delta_{ij} \frac{d \phi ^{j}}{dr} \right)  +
\phi^{*i}   \left(  M_{ij}-\frac{\omega^2 { H_i^2 } r}{F}\delta_{ij} \right)\phi^{j} , 
\end{eqnarray}
where the matrix $M_{ij}$ is given by 
\begin{eqnarray}
\label{mimat}
M_{ij} =  \frac{4 r_+^6}{L^2 r^5 {\cal H} }( 1+k_i) m_i (1+k_{j}) m_{j} . 
\end{eqnarray}
In  (\ref{varact2}) summation over the indices $i, j$ is implied. 
It is easy to see that on variation of the action  in  (\ref{varact2}) by 
$\phi^{i *}$  we obtain the equations given in (\ref{gageqnr}).

Finally we will describe the behaviour of the solutions to the equations in (\ref{varact2}) 
near the horizon and the boundary. 
Near the horizon $r_+$ the equations decouple and the fluctuations $\phi^i$  
 behave as 
\begin{equation}
\label{nrhora}
\phi^i (r) \sim ( r - r_+) ^ {\pm i  \alpha \omega} , 
\end{equation}
where $\alpha$  is given by 
\begin{eqnarray}
 \alpha &=& \frac{1}{F_h} \sqrt{( 1+k_1)(1+k_2)(1+k_3)} \\ \nonumber
  F &=& ( r-r_+) F_h + \cdots , \qquad 
F_h =  \frac{2 r_+}{L^2} ( 2 + k_1 + k_2 + k_3 - k_1k_2k_3). 
\end{eqnarray}
Near the boundary $r\rightarrow \infty$ the geometry reduces to 
that of $AdS_4$. From the equation (\ref{gageqnr}) it is easy the fluctuations  decouple 
at $r\rightarrow \infty$. The behaviour of the two independent 
solutions of each of the fluctuations in the limit $r\rightarrow\infty$  is given by 
\begin{eqnarray}
\label{asmsol}
\phi^i(r) \sim \frac{1}{i r L^2 \omega}I_1(\frac{i L^2\omega}{r})\sim r^{-2}, \qquad 
\phi^i(r)\sim \frac{1}{i r L^2 \omega}K_1(\frac{i L^2\omega}{r})\sim {\rm{constant}}. 
\end{eqnarray}

Before we begin our analysis we will review the procedure to obtain the retarded Green's
of the R-Currents given by 
\begin{equation}
 G^{i} (\omega,  q ) = i \int dt d^3 x e^{i (\omega t - q x)}
 \theta(t) \langle [J_x^i (t, \vec x), J_x^i ] \rangle, 
\end{equation}
where the expectation value is taken in the thermal state at finite chemical potential. 
We will restrict our attention to the Green's function with diagonal entries in the 
R-symmetry index.  To obtain this Green's function from the gravity dual 
we will follow the procedure developed by \cite{Policastro:2002se,Kovtun:2005ev}. 
We will need to solve the coupled  set of equations in (\ref{gageqnr}) with ingoing 
boundary conditions at the horizon which is given by 
\begin{equation}
\label{nrhor}
\phi^i (r) \sim ( r - r_+) ^ {- i  \alpha \omega} , 
\end{equation}
With this ingoing boundary conditions at the horizon we obtain the 
solution at the boundary $r\rightarrow\infty$. Once this is done,   the 
retarded Green's function in Fourier space is given by 
\begin{eqnarray}
 G_T^i (\omega ) & =&  \hat G^i(\omega, T )  + G_{ \rm{counter}} ( \omega, T ), \\ \nonumber
\hat G^i(\omega, T ) &=& - 
\left. \frac{N^2}{8\pi^2 L^3}  \lim_{r\rightarrow \infty} \frac{r F \phi^{i\prime}}{L  \phi^i} \right|_{\phi^j_\infty=0, j\neq i},  
\end{eqnarray}
where we have dropped  the momentum dependence since we are interested in $q=0$.
We have also explicitly denoted the temperature dependence by $T$ in the Green's function.  
 $G_{\rm counter}(\omega, T) $ is  the counter term needed  to cancel
the $\log(r)$ divergences. 
As it will be clear in our subsequent discussion 
we will not be requiring the explicit form of these terms. 
Thus the important properties of the Green's function is essentially contained in the 
function
\begin{equation}
\label{grdef}
 g^{i} (\omega) =  \lim_{r\rightarrow \infty} \left. \frac{r F }{L \phi^i} \frac{d \phi^i}{dr}
\right|_{\phi^j_\infty =0 , j\neq i} 
\end{equation}
Note that since we are solving the coupled set of equations in (\ref{gageqnr}) 
with the boundary conditions
(\ref{nrhor}), there will be $3$ independent constants. These are the values of $\phi^i$ at the boundary
denoted by $\phi^i_\infty$. The $i$-th  diagonal component of the correlator can be obtained 
by  first evaluating the ratio given in (\ref{grdef}) and then setting the boundary values 
of $\phi^j_\infty =0$ for $j\neq i$.  A simple way to understand this condition is  that the boundary 
effective action  from which 
the Green's function at frequency $\omega$  is evaluated  will be a quadratic functional of $\phi^i_\infty$
\begin{equation}
S_{ {\rm Boundary }} = \sum_{i, j=1}^3 \phi^i_\infty \ G^{ij}(\omega)  \phi^j_\infty. 
\end{equation}
Then to obtain the diagonal component $G^{ii}$ we can examine the response in 
$\phi^i$ by setting all the other sources to $\phi^j, j\neq i$ to zero. 
Thus  to study the behaviour of the retarded Green's function in the frequency 
plane, it is sufficient 
to study the function $g^i(\omega)$.

\subsection{Green's  function in the $\omega$-plane}

In this subsection we  will discuss the analytic properties  of the 
Greens function $g^{i}$ in the complex $\omega$ plane. 
We will show that this function does not have any poles or branch cuts in the 
upper half $\omega$  plane. We then will obtain the asymptotic behaviour
of this function in the $\omega\rightarrow i \infty$ limit. 
The analysis in this section is an extension of the one performed
in \cite{David:2011hy} for shear correlator. For the shear correlator the behaviour
of the Green's function is determined by a single differential equation. 
For the situation at hand 
we need to examine the properties of the  coupled 
set of equations given in (\ref{gageqnr}).

\vspace{.5cm}
\noindent
{\bf  No poles for ${\rm Im} \, \omega >0$ }
\vspace{.5cm}

It is known  that 
poles or divergences  in the retarded Green's function of an operator  corresponds to the 
the quasi-normal modes of the differential equation satisfied by the 
corresponding dual field \cite{Kovtun:2005ev}.   In fact 
by examining the definition of $g^i$ given in (\ref{grdef}) we see that 
poles will correspond to the situation when $\phi^i_\infty$ also vanish. 
Thus poles in 
$g^{i} (\omega)$ corresponds to quasi-normal modes of the equation 
(\ref{gageqnr}).  Quasi-normal modes are solutions to the equation 
(\ref{gageqnr}) with the following boundary conditions
\begin{eqnarray}
\label{bcgag}
 & & \phi^i(r) \sim ( r-r_+)e^{-i\alpha \omega}, \qquad r\rightarrow r_+, \\ \nonumber
 & & \phi^i(r) \sim r^{-2}, \qquad r\rightarrow \infty.
\end{eqnarray}
for all $i$. 
We will now show that such quasinormal modes do not exist with ${\rm Im} \, \omega >0$
by examining the equation (\ref{gageqnr}) and the action (\ref{varact2}). 
Note that all the coefficients in the equation (\ref{gageqnr}) are real. 
Therefore if $\phi^i(r)$ is a quasinormal mode with frequency 
$\omega$ then $(  \phi^i(r) )^*$ is a quasinormal mode with 
frequency $\omega^*$.  This is easily demonstrated by 
taking the complex conjugate of the  the set of equations in (\ref{gageqnr}). 
Let us consider the equation $S_\phi - S_\phi=0$. By substituting the solutions
 $\phi^i(r)$ and $(  \phi^i(r) )^*$ into this equation and integration 
 by parts and then using the equations of motion given in (\ref{gageqnr}).
we obtain
\begin{equation}
 0 = \sum_{i=1}^3
\left( Fr\f{H_i^2}{\cal H} ( \phi^{i^*\prime}\phi^{i} -\phi^{i^\prime}\phi^{i*} ) |_{r_h}^\infty
+ ( \omega^{*2} - \omega^2 ) \int_{r_h}^\infty dr \frac{ {\cal H} r}{ F} \phi^i \phi^{i*}  
\right). 
\end{equation}
Here we have used the equations of motion of $(\phi^i) *$ in the first $S_{\phi}$ and 
the equation of motion of $\phi^i$ in the second $S_{\phi}$. 
We have also used the fact that the matrix $M_{ij}$ in (\ref{mimat}) is symmetric. 
From the conditions for the quasi-normal modes given in 
(\ref{bcgag})  and the fact that 
 $ F \sim ( r- r_h) F_h$ as $ r\rightarrow r_h$, 
we see that the first term in the above equation vanishes 
for ${\rm{Im}}( \omega ) >0$.  Since the second term is 
positive definite, we  are  led to the situation
\begin{equation}
 \omega^2 =  \omega^{*2}.
\end{equation}
since the second term is positive definite.
Thus we see that $\omega$ has to be either purely real or purely imaginary.
This together  
with the condition ${\rm{Im}}( \omega ) >0$ results in the fact that
$\omega$ is restricted to lie  on the upper imaginary axis. 

We now assume that there exists a quasinormal mode on the upper imaginary 
axis with $\omega^2<0$. Then $S_\phi$ is given by 
\begin{eqnarray}
\label{varact22}
 S_{\phi}= \int_{r_h}^{\infty}dr \frac{F r H_i^2}{{\cal H}}
\left(  \frac{d \phi^{i *}}{dr} \delta_{ij} \frac{d \phi^{j}}{dr} \right)  +
\phi^{i*}   \left(  M_{ij}+\frac{|\omega|^2 { H_i^2 } r}{F}\delta_{ij} \right)\phi^{j}. 
\end{eqnarray}
The first and third  term in the action are manifestly positive definite. 
The  only term one needs to check is the following expression
\begin{equation}
\label{defi}
 I_\phi= \int_{r_h}^\infty dr \phi^{i*}M_{ij} \phi^j, 
\end{equation}
where the matrix is given in (\ref{mimat}). Since the matrix
$M_{ij}$ is symmetric, to  prove this term is positive 
definite,  it is sufficient to show that the eigen values of this matrix is 
greater than equal to zero.  From the form of the matrix $M_{ij}$ it 
is easy to show that the eigen values are given by 
\begin{equation}
 \left(  \frac{ 4 r_+^6}{L^2 r^5 {\cal H}} \sum_{i=1}^3  ( 1+k_i)^2 m_i^2  , \;\;0, \;\;0 \right).  
\end{equation}
Thus   the term in  (\ref{defi}) is positive definite. Therefore the 
action  $S_\phi$ for $\omega^2 <0$ is positive.  Now substituting the quasinormal 
mode with $\omega^2<0$  into  $S_a$ and integrating by parts 
we obtain
\begin{equation}
 S_{\phi} = \sum_{i=1}^3 \frac{F r H_i^2}{{\cal H}} 
\phi^{i\prime}\phi^{ i*} |_{r_h}^\infty.
\end{equation}
From the boundary conditions of  $\phi^i(r)$ for a quasinormal mode given in (\ref{bcgag})
and from the condition ${\rm Im} (\omega ) >0$ we see that the boundary terms given 
above vanish. Thus we obtain $S_{\phi}=0$, but since $S_\phi$ is positive definite 
we must have $\phi^i=0$. Thus no quasinormal modes exist in the 
upper half plane which implies no poles or divergences exist in the Green's
function in this domain. 

\vspace{.5cm}
\noindent
{\bf No poles for  $\omega$ real and  $\omega \neq 0$}
\vspace{.5cm}

To show that there are no poles for $\omega$ real and  $\omega \neq 0$ we will first 
redefine the variables
\begin{equation}
\label{redefd}
\phi^i =  \Phi_i(r) \varphi^i , \qquad \Phi_i (r) = \sqrt{\f{{\cal H}}{r F H_i^2}}.  
\end{equation}
Note that  here there is no summation over the index $i$. 
Substituting this redefinition in the equation (\ref{gageqnr}) we obtain
the following equation 
\begin{equation}
\label{redefgeq}
\varphi^{i\prime\prime}  + \left( \frac{ \Phi_i''}{\Phi_i} + 
\left( \ln ( \frac{r F H_i^2}{{\cal H}} ) \right)^\prime \frac{\Phi_i'}{\Phi_i} \right) \varphi^{i \prime}
+ \omega^2 \frac{\cal H}{F^2}\varphi^i - \tilde M_{ij} \varphi^j =0, 
\end{equation}
where
\begin{equation}
\tilde M_{ij} = \frac{4 r_+^6}{ r^6 L^2 F} ( 1+k_i) \frac{m_i}{H_i} ( 1+k_j) \frac{m_j}{H_j}.  
\end{equation}
If  $\omega$ is real, then $(\varphi^i)*$ is also another solution of the equations in 
(\ref{redefgeq}). Now consider the quantity
\begin{equation}
W(r)  = \sum_{i=1}^3 \left( 
( \varphi^{i\prime} (\varphi^i)* -  (\varphi^i)^{*\prime} \varphi^i \right). 
\end{equation}
Using the equations in (\ref{redefgeq}) and the fact that $\tilde M_{ij}$ is symmetric 
and real it is easy to see that $W$ satisfies the equation
\begin{equation}
W'(r) = 0, \qquad W(r) = C, 
\end{equation}
where $C$ is a constant. 
Now let us  assume that a quasinormal mode with $\omega$ real and  $\omega \neq 0$ 
exists. 
We can then determine the value of the constant by examining the boundary conditions 
for the quasinormal mode. In the limit $r\rightarrow r_+$ using (\ref{bcgag}) and
the redefinition (\ref{redefd}) 
we obtain
\begin{equation}
\varphi^i \rightarrow 
\varphi^i_{r_+}( r-r_+) ^{ -i\alpha\omega + \frac{1}{2}}, \qquad r \rightarrow r_+, 
\end{equation}
where $\varphi^i_{r_+}$ are constants. 
Substituting this in the definition of  $W$ we get
\begin{equation}
\label{valW}
W = -2i \omega \alpha \sum_{i=1}^3|\varphi^i_{r_+} |^2. 
\end{equation}
Now lets examine $W$ near the boundary. 
Again from the equations (\ref{bcgag}) and (\ref{redefd}) 
we see that
\begin{equation}
\varphi^i \rightarrow r^{-1/2}, \qquad r\rightarrow \infty. 
\end{equation}
Using this in the equation for $W$ we see that $W\rightarrow \frac{1}{r^2}$ as $r\rightarrow \infty$. 
This contradicts our conclusion in (\ref{valW}). 
Thus the quasi-normal modes and hence poles of the Green's function do not exist 
for $\omega$ real and  $\omega \neq 0$. 

\vspace{.5cm}
\noindent
{\bf No poles for  $\omega=0$}
\vspace{.5cm}

For this analysis it is convenient to rewrite the equations in  (\ref{gageqnr}) as 
\begin{eqnarray}
\label{gageqnr1}
\phi^{''i}+\left( \ln(\frac{fH_i^2}{H})'+\f{3}{r} \right)\phi^{i\prime}+\frac{L^4\omega^2{\cal H} }{r^4 f^2}\phi^{i}-
\f{4 r_+^6}{r^8}\prod_{j=1}^3(1+k_j)\sqrt{k_i}\sum_{l}\left(  \frac{\sqrt{k_l}}{H_i^2 f}  
\phi^{l}\right)  &=&0. \nonumber\\
\end{eqnarray}
Here we have rewritten $m_i$ and $F$ in terms of their original definitions. 
For $\omega=0$ the above equations can be exactly solved  \cite{Jain:2009pw}. 
The solution which is finite at $r\rightarrow r_+$ is given by 
\begin{eqnarray}
\label{deftau}
\phi^i&=& \tau_{ij}\phi^j_{\infty} , \qquad  \tau_{ij}\equiv\delta_{ij}-\f{1}{2H_i}\sqrt{k_i k_j}u
\end{eqnarray}
where $\phi^i_{\infty} $ are the values of $\phi^i$ at the boundary.  
We will now  determine the behaviour of the Greens's function in the limit $\omega\rightarrow 0$. 
Let the solution for the equation (\ref{gageqnr}) for finite $\o$ be 
\bea
\label{ansol}
\phi^i (r)= \T_{ij}\psi^j (r). 
\eea
Substituting (\ref{ansol}) in (\ref{gageqnr}) we have
\bea
\label{psieq}
\T_{ij}\psi^{''j}+2\T^{'}_{ij}\psi'^{j}+\left( \ln{\f{fr^3H_i^2}{{\cal H}}}
\right)'\T_{ij}\psi^{'j}+\f{\o^2 {\cal H }L^4}{r^4 f^2}\T_{ij}\psi^j=0. 
\eea
From the definition in (\ref{ansol}) and   (\ref{deftau}) we see that the boundary values 
$\phi_{\infty}^i $ are related to the  boundary values 
$\psi^i_\infty$ by 
\begin{equation}
\psi^i_{\infty} = \phi_{\infty}^i. 
\end{equation}
Then from the equation (\ref{psieq}) we see that in the $\omega\rightarrow 0$ limit, the 
solution is given by  the constants
\begin{equation}
\label{psilim}
\psi^i(r)_{\omega\rightarrow 0} = \psi^i_\infty = \phi_{\infty}^i. 
\end{equation}
Since the $\psi^i$ in the $\omega\rightarrow 0$ limit are constants, 
we see that the values of $\phi^i$ are the horizon in the $\omega\rightarrow 0$ limit
are given by 
\begin{equation}
\phi^i(r\rightarrow r_+)_{\omega\rightarrow 0} = \tau_{ij}^{}(r_+)  \psi^j_\infty= 
\tau_{ij}^{}(r_+)  \phi^j_\infty. 
\end{equation}
Now let us obtain the equation for the derivative 
\begin{equation}
f_i = \tau_{ij} \psi^{j\prime} (r). 
\end{equation}
Using the differential equation (\ref{psieq}) we see that $f^i$ satisfy the equation
\begin{equation}
\label{eqef}
f_i' + \tau_{ik}' \tau_{kj}^{-1} f_j + \left( \ln{\f{fr^3H_i^2}{{\cal H}}}\right)' f_i + \f{\o^2 {\cal H }L^4}{r^4 f^2}\T_{ij}\psi^j=0. 
\end{equation}
The equation (\ref{psilim}) implies at in the $\omega\rightarrow 0$ limit $f^i \rightarrow 0$. 
We are interested in the behaviour of $f^i$ to the linear order in $\omega$. 
From (\ref{eqef}) we see that to this order it is sufficient to solve the equation
\begin{equation}
\label{eqef1}
f_i' + \tau_{ik}' \tau_{kj}^{-1} f^j + \left( \ln{\f{fr^3H_i^2}{{\cal H}}}\right)' f_i =0. 
\end{equation}
The initial conditions of this equation at the horizon can be obtained from the ingoing 
boundary conditions for $\phi^i$ which translates to ingoing 
boundary conditions for $\psi^i$.  This is given by 
\begin{equation}
f_i(r \rightarrow r_+) = ( -i \omega \alpha) \tau_{ij} (r_+) ( r-r_+)^{-i\alpha \omega-1} \psi^j(r_+). 
\end{equation}
Thus in the $\omega\rightarrow 0$ limit we see that the $f^i$ has the following behaviour 
at the horizon
\begin{equation}
\label{inicond}
f_i(r \rightarrow r_+)_{\omega\rightarrow 0} = ( -i \omega \alpha) \tau_{ij} (r_+) 
( r-r_+)^{ -1} \phi^j(\infty), 
\end{equation}
where we have used the fact that in the $\omega\rightarrow 0$ limit the 
$\psi^i$'s are constants given by (\ref{psilim}). 
We now need to solve  (\ref{eqef1}) such that it obeys the initial conditions
(\ref{inicond}). 
This solution is given by 
\begin{equation} \label{solff}
f_{i} = \frac{{\cal H}}{f r^3 H_i^2} (\tau^{T})^{-1}_{ij} f_j^{r_+}, 
\end{equation}
where
\begin{equation}
(\tau^{T})^{-1}_{ij} = \delta_{ij} + \frac{\sqrt{k_ik_j} u }{2 H_j (  1- \sum_{i=1}^3 \frac{k_i u}{ 2 H_i} ) }, 
\end{equation}
and $f_j^{r_+}$ are determined by the initial conditions given in (\ref{inicond}). 
This is given by 
\begin{equation}
f_j^{r_+} = - \frac{i\omega r_+L^2}{\sqrt{( 1+ k_1) ( 1+ k_2) ( 1+k_3) } }
 \sum_{l =1}^3\tau^{T}_{jl }( r_+) ( 1+ k_l)^2  \tau_{lm}(r_+)  \phi^m_\infty . 
\end{equation}
Now that we have the solution for $f_j$, we can obtain the leading 
behaviour of the Green's function in the $\omega\rightarrow 0$ limit. 
\begin{eqnarray} \label{solgd}
g^{i} |_{\omega\rightarrow 0}
&=&  \left.  \lim_{r\rightarrow \infty} \frac{r F }{L \phi^i} \frac{d \phi^i}{dr}\right|_{\phi^j_\infty =0, j\neq i},   \\ \nonumber
&=&  \lim_{r\rightarrow\infty}\frac{r^3}{L^3} \tau_{ii}^{\prime} 
+\left.  \frac{ f_i^{r_+} }{L^3 \phi_i^\infty} \right|_{\phi^j_\infty =0, j\neq i}, \\  \nonumber
&=& \frac{k_i r_+^2}{L^3 }  -
\frac{i\omega r_+}{4L  \sqrt{( 1+ k_1) ( 1+ k_2) ( 1+k_3) }} ( 4 + 4 k_i  + k_i \sum_{j=1}^3 ( k_j) ). 
\end{eqnarray}
In the second line we have used the definition in (\ref{ansol}) and the solution for $f_i$ given in 
(\ref{solff}). 
The same  analysis can easily be extended  to obtain the full retarded correlator in the 
$\omega\rightarrow 0$ limit. This results in 
\begin{equation}\label{solmatg}
 g^{ij}|_{\omega\rightarrow 0} = 
\frac{ \sqrt{k_ik_j} r_+^2}{L^3} - \frac{ i\omega r_+}{ \sqrt{( 1+ k_1) ( 1+ k_2) ( 1 + k_3) }} 
\sum_{l=1}^3 \tau^{T}_{il}(r_+) ( 1+ k_l)^2 \tau_{lj} . 
\end{equation}
From this it is easy to evaluate the real and imaginary part of the conductivity matrix at 
zero frequency. This is given by 
\begin{equation}\label{solmatcon}
 \sigma^{ij}_{\omega\rightarrow 0} =- 
\frac{N^2}{8\pi^2 L^3}  \lim_{\omega\rightarrow 0} \frac{1}{i \omega} g^{ij}(\omega) , 
\end{equation}
where $g^{ij}$ in the limit $\omega\rightarrow0$ is given in (\ref{solmatg}). 
The real part of the 
 conductivity matrix has been evaluated earlier by  \cite{Jain:2009pw} using a slightly 
different approach.  We have verified that the 
real part of the conductivity given in (\ref{solmatcon}) agrees with the 
expression given below equation (4.23) of \cite{Jain:2009pw}. 
account the fact that the gauge fields in this paper are dimensionless.
Note that the imaginary part of the conductivity has a pole at $\omega\rightarrow 0$. 
The residue of the pole is the constant value of the Green's function in the $\omega\rightarrow 0$ limit.
For future reference we will write this equation as 
\begin{eqnarray} \label{imgreen}
 \lim_{\omega \rightarrow 0} \omega {\rm Im} \sigma^{i} &=&  
\frac{1}{e^2} 
 g^i|_{\omega\rightarrow 0}
=  \frac{1}{e^2} \frac{k_i r_+^2}{ L^3}., \\ \nonumber
\frac{1}{e^2 }&=& \frac{N^2}{ 8\pi^2 L^3}  
\end{eqnarray}
Here we have just written the relation for the diagonal 
entries in the Green's function.  
Thus the above analysis which resulted in the equation (\ref{solgd}) and (\ref{solmatg}) 
shows that the 
Greens' function $g^i, g^{ij}$ has a well defined  behaviour in the 
$\omega\rightarrow 0$ limit. 

We can now show that the Green's 
function admits a systematic power series expansion around the origin.  
Let define
\begin{equation}
\tilde g _{ik} (r) =  \frac{\tau_{ij}\psi^{j\prime}}{\omega \psi^k} = \frac{f_i}{\omega \psi^k}. 
\end{equation}
From our analysis of the solutions $f_i$ and $\phi^i$ in the limit $\omega\rightarrow 0$ we see
that $\tilde g_{ik}$  has a well defined in the limit $\omega\rightarrow 0 $. 
This is because $f_i$ behaves linearly for small $\omega$.   
Using  (\ref{psieq}) we see that  $\tilde g_{ik}$ satisfies the following equation
\begin{eqnarray}\label{nonlin}
\tilde g_{ik}' +  \tau_{ij} \tau^{-1}_{jl} \tilde g_{lk} 
 +  \left( 
\ln{\f{fH_i^2}{{\cal H}}} \right)' \tilde g_{ik}  + \frac{\omega {\cal H}}{3  F^2} \tau_{ij} \hat g_{il} g_{lk}
+ \omega \tilde g_{ik} \tau^{-1}_{k j} \tilde g_{jk} =0, 
\end{eqnarray}
where
\begin{equation}
\hat g_{ki} = \frac{1}{\tilde g_{ik} }.
\end{equation}
In (\ref{nonlin}) the indices $i, k$ are free indices and are not summed over,  all the other
repeated indices are summed over. 
Note that the equation (\ref{nonlin}) 
is a non-linear equation for $\tilde g_{ik}$ and it admits a
power series expansion in $\omega$. 
Since the solution is known in the $\omega\rightarrow 0$ limit, we can set up 
a power series expansion around this solution. 
From  $\tilde g_{ik}$,  it is easy to obtain an expansion of the Green's function
as a power series in $\omega$.  The Green's function is given by 
\begin{eqnarray}
g^{i}
&=&  \left.  \lim_{r\rightarrow \infty} \frac{r F }{L \phi^i} \frac{d \phi^i}{dr} 
\right|_{\phi^j_\infty =0, j\neq i},
\\ \nonumber
&=& 
  \lim_{r\rightarrow \infty} \frac{r F }{L  }\left( \tau^{\prime}_{ii} +
  \omega \tilde g_{ii}|_{\phi^j_\infty =0, j\neq i}
 \right). 
\end{eqnarray}
To conclude we have shown that the Green's function of interest has a well defined limit 
as $\omega\rightarrow 0$ and it admits a power series in $\omega$ at the origin
in the complex $\omega$ plane.

\vspace{.5cm}
\noindent
{\bf Absence of branch cuts for ${\rm Im}\,  \omega \geq 0$}
\vspace{.5cm}

The argument for the absence of branch cuts is essentially same as the one 
developed in \cite{David:2011hy} for the shear tensor correlator. 
The  important properties of the Green's function is determined by the 
function defined in (\ref{grdef}). This  is determined by solving the differential equation
in (\ref{gageqnr}) with the ingoing boundary condition at the horizon (\ref{nrhor}).
Both the differential equation and the boundary conditions are smooth with respect to 
$\omega$.  
Now   the theory of ordinary differential equations ensures that a local 
Forbenius expansion of the solution is smooth with respect to the parameters of the 
differential equation provided the equation and the boundary conditions are smooth with 
respect to the parameters \cite{Arnold}. 
Thus  $\phi^i$ and  its radial derivative at the boundary must be smooth 
with respect to $\omega$. Then from the definition of the 
Green's function (\ref{grdef}), the only locations at which it or  its 
$n$-th order derivative with respect to $\omega$ can be singular 
is when $\phi^i$ satisfies the quasinormal mode boundary conditions given in 
(\ref{bcgag}).  This is because the denominator in (\ref{grdef}) vanishes for 
the quasi-normal mode boundary conditions. 
But as we have already argued that these modes do not occur   in the
upper half plane. Thus we have the result that the Green's function is smooth 
with respect to $\omega$ in the upper $\omega$-plane.

\vspace{.5cm}\noindent
{\bf  Behaviour as $\omega \rightarrow \infty$}
\vspace{.5cm}

To obtain the behaviour of the Green's function at large $\omega$ we 
follow the procedure developed in \cite{David:2011hy}. We first 
rewrite the differential equation given in (\ref{gageqnr}) by defining 
a dimensionless variables 
\begin{equation}
 z = \frac{r_+}{r},  \qquad  i  \lambda =  \frac{L^2}{r_+} \omega . 
\end{equation}
\begin{eqnarray}
\label{difcz}
\phi^{i \prime\prime}+\left( \ln  \frac{fH_i^2}{z {\cal H}}  \right) '{\phi}^{i\prime}-
\frac{\l^2 {\cal H} }{f^2}{\phi}^{i \prime}-
(1+k_i)m_i\sum_{j}\left(  \frac{4z^4}{H_i^2f^2}(1+k_{j}) m_{j} {\phi}^{j} \right) =0. \nonumber
\\
\end{eqnarray}
For convenience we have gone over to the Euclidean frequency labelled as $\l$.
We are interested in obtaining  the behaviour 
of the functions $\phi^i$  as $\l\rightarrow \infty$. 
For this purpose we rescale the co-ordinates as 
\begin{eqnarray}
 y=\l z.
\end{eqnarray}
This leads to the following equations for the gauge field fluctuations as a $1/\lambda$ expansion.
\begin{eqnarray}
\label{llameq}
 {\phi}^{i \prime\prime }+ \left(  \frac{f'}{f} -\frac{1}{y} + \frac{2y}{\l^2} J_i 
 \right) {\phi}^{i\prime} - 
\frac{2y^4}{H_i^2 \l^6 f}(1+k_i)m_i\sum_j(1+k_j)m_j {\phi}^j - \frac{H}{f^2}{\phi}^i=0.\nonumber\\
\end{eqnarray}
where
\begin{equation}
J_1 = (\frac{k_1}{H_1}-\frac{k_2}{H_2}-\frac{k_3}{H_3}), 
\quad
J_2 = (\frac{k_2}{H_2}-\frac{k_3}{H_3}-\frac{k_1}{H_1}),\quad
J_3= (\frac{k_3}{H_3}-\frac{k_1}{H_1}-\frac{k_2}{H_2}),
\end{equation}
and there is no sum over $i$ in (\ref{llameq}). 
In terms of the dimensionless variables the in going boundary conditions 
(\ref{nrhor}) can be re-written as 
\begin{equation}
\label{nrhorl}
\phi^i \sim \left( 1- \frac{y}{\lambda} 
\right)^{ \frac{\lambda r_+ \alpha}{L^2} }, \qquad y \rightarrow \lambda.
\end{equation}
We can set up a Forbeinus expansion of the solutions to the set of equations in 
(\ref{llameq}) obeying the boundary conditions (\ref{nrhorl}). 
This expansion is valid in the domain $y\sim \lambda$. 
One can also set up an expansion for $y\rightarrow 0$ which can be organized as 
a systematic expansion in powers of $1/\lambda$. 
It is important to note when  $\lambda$ is  strictly infinity the equations in 
(\ref{llameq}) decouples and reduces to that of three vector fields in the 
background of pure $AdS_5$. 
Thus the leading solutions  to the equations  in the large $\lambda$ expansion will be identical
to the zero temperature case. 
We will construct the $6$ independent solutions of the coupled equations 
given in (\ref{llameq}) as an expansion in $1/\lambda$ as follows. 
Let us define
\bea \label{abc}
a(y)=\f{ {\phi}^{1\prime}(y)}{\phi^1 (y)},  \qquad
b(y)=\f{{\phi}^{2\prime}(y)}{\phi^2 (y)},  \qquad
c(y)=\f{{\phi}^{3\prime}(y)  }{\phi^3 (y)}, 
\eea
where the derivatives are   with respect to $y$.
We will now focus on obtaining a perturbative expansion for  $a(y)$,  a similar analysis can be 
performed for the quantities $b(y)$, $c(y)$. 
From (\ref{llameq}) we can obtain  
the equation for $a(y)$ which can be given as  
\bea
\label{redeqn}
& &  a^{\prime} +(a^{})^2 + \left\{  \frac{f'}{f} -\frac{1}{y} +
 \frac{2y}{\l^2}\left( \frac{k_1}{H_1}-\frac{k_2}{H_2}-\frac{k_3}{H_3}\right)\right\}a \\ \nonumber
 & &\qquad \qquad  - 
\frac{4y^4}{H_1^2 \l^6 f}(1+k_1)m_1\sum_j(1+k_j)m_j \frac{{\phi}^j}{{\phi}^1} - \frac{H}{f^2}=0.
\eea
From the equation (\ref{redeqn}) we see that the $a$ admits an expansion of the form 
\begin{equation}
 a = a_0 +\frac{1}{\l^2} a_1 +\frac{1}{\l^4} a_2 +{\cal{O}}(\frac{1}{\l^6}). 
\end{equation}
By substituting this expansion for $a$ into (\ref{redeqn}) and matching 
orders in $\frac{1}{\lambda^2}$ we can obtain equations which determine the functions $a_{i}$.
The leading orders in the expansion are determined by the following equations
\bea
\label{expad4}
& & a_0^{\prime }+(a_0^{ })^2-\f{1}{y}a_0^{ }-1=0, \\
& & a^{'}_1 + (2 a_0-\frac{1}{y})a_1+4 k_1y a_0+(k_1+k_2+k_3)y^2 =0,\nn\\
& & a^{'}_2 + (2 a_0-\frac{1}{y})a^1_2+(a_1)^{2} +\left((1+k_1)(1+k_2)(1+k_3)+k_1^2 \right)y^3 a_0+4k_1ya_1\nonumber\\
& &-  \left( 2 (1+k_1)(1+k_2)(1+k_3) + k_1^2 + k_2 ^2 + k_3^2  \right) y^4 
=0.\nonumber
\eea
The two independent solutions for the first equation in (\ref{expad4}) are 
\bea
\label{indsol}
{a_0}^{(1)} = -\frac{K_{0}(y)}{K_{1}(y) }= \frac{d}{dy} \ln(y K_1(y)), \qquad 
{a_0}^{(2)} =   \frac{I_{0}(y)}{I_{1}(y)}= \frac{d}{dy} \ln(y I_1(y)).  
\eea
We now show how to obtain a systematic expansion around the first solution $a_0^{(1)}$. 
The first order correction is given by 
\begin{equation}
\label{b1soln}
 {a_1}^{(1)} = -2k_1y-\left(\f{k_1+k_2+k_3} {6} \right) y^3(1-\f{K_2^2}{K_1^2})+\f{c_1}{yK_1^2}.
\end{equation}
We set $c_1=0,$ so as  not to change the asymptotics of this solution at $y\rightarrow\infty$.
Substituting this into the equation for  $a_2^{(1)}$ it is easy to obtain the solution. 
However for us it is sufficient to obtain the behaviour near $y\rightarrow 0$. 
This is given by 
\begin{equation}
\label{b2sol}
 {a_2}^{(1)}(y) =  y^3 + O(y^5, y^5 \log y). 
\end{equation}
A similar expansion can be set up for the functions $b$ and $c$ defined in (\ref{abc}). 
Examining the equations for $a_n$ it is easy to see that near the origin their behaviour is 
given by 
\begin{equation}
\label{bnsol}
 a_n^{(1)}(y) \sim  y^m, \qquad m \geq 3, \qquad \hbox{for}\quad n\geq 2
\end{equation}
Note that the equation which determines $a$ couples with the functions $b$ and $c$ at 
$O(\frac{1}{\lambda^6})$ and therefore the coupling can be treated perturbatively. 
Thus a systematic expansion for the first solution to $a$ can be found. 
This is given by 
\begin{eqnarray}
\label{chphi1}
 \phi^{1}_{(1)}(y) &=& \exp\left[ \int_0^y dy \left(a_0 + 
\frac{ {a_1}^{(1)} } {\lambda^2}  + \frac{ {a_2}^{(1)} }{\lambda^4}
+ \cdots \right) \right], \\ \nonumber
&=& y K_1(y) \left( 1 + \frac{1}{\lambda^2} \int_0^y dy a_1^{(1)} (y)   + \cdots \right).
\end{eqnarray}
Note that 
\begin{equation}
{{\phi}^1}_{(1)} (y) \sim {\rm constant}, \quad y \rightarrow 0.
\end{equation}
From (\ref{b1soln}) we see that the  $O( 1/\l^2 )$ term in 
(\ref{chphi1})  goes as $y^2$  near the origin. The higher order terms (\ref{b2sol})are 
further suppressed as $y\rightarrow0$. 
We now obtain the second solution   starting with the 
zeroth order second solution given in (\ref{indsol}). 
The first order correction about the second solution is given by 
 \begin{eqnarray}
\label{b1soln2}
 {a_1}^{(2)} &=& -2k_1y+(\f{k_1+k_2+k_3)}{6}y^3(1-\f{I_2^2}{I_1^2}),\nonumber\\
&=& -2k_1y+\f{k_1+k_2+k_3)}{6}(y^3 - \f{y^5}{16})+{\cal O}(y^7).
\end{eqnarray}
Similarly it can be shown that  ${a_2}^{(2)}(y)$  has the following 
behaviour 
\bea \label{b2soln2}
a_2^{(2)}(y)\sim y^3+{\cal O}(y^5).
\eea
Thus we can write the second  solution as 
\begin{eqnarray}
\label{chphi2}
 {{\phi}^1 }_{(2)}(y) &=& \exp\left[  \int_0^y dy \left(a^{(2)}_0 + 
\frac{a^{(2)}_1}{\lambda^2}  
+ \cdots \right) \right], \\ \nonumber
&=& y I_1(y) \left( 1 + \frac{1}{\lambda^2} \int_0^y dya^{(2)}_1 (y)    + \cdots \right).
\end{eqnarray}
Note that  the leading behaviour of the second solution  at the origin is given by  
\begin{eqnarray}
\label{b0soln2}
 {{\phi}^1 }_{(2)}(y)\sim y^2 , \qquad y\rightarrow0. 
\end{eqnarray}
In this way one can obtain the $6$ independent solutions to the coupled set of equations
given in (\ref{gageqnr}).  These solutions are all valid as expansions around the origin $y=0$. 

Thus the  general solution for the $\phi^1$ is the linear combination 
\bea
\label{matsol}
\phi^1(y) = \alpha(\lambda){{\phi}^1 }_{(1)}(y)+ \beta(\lambda){{\phi}^1 }_{(2)}(y). 
\eea
Similar solutions can be written for $\phi^2, \phi^3$. 
The coefficients  $\alpha(\lambda), \beta(\lambda)$  are determined by extrapolating
the Forbeinus  expansions near the horizon $y\sim \lambda$ to $y\rightarrow 0$ and 
matching with the expansions in (\ref{matsol}) extrapolated to $y\rightarrow \infty$. 
For our purpose we only  need the large $\lambda$ behaviour of these coefficients. 
We see that 
\begin{equation}
\label{basymch} 
\beta(\lambda)  \rightarrow 0, \qquad 
\alpha(\lambda) \rightarrow 1, \qquad \lambda  \rightarrow \infty.
\end{equation}
The reason for this is that 
when $\lambda$  is strictly $\infty$, the equation (\ref{llameq}) reduces to that 
of the zero temperature case  as we have observed earlier. In  this situation
the solution which is finite as $y\rightarrow \infty$  is given by $y K_1(y)$ \footnote{We have chosen
the normalization of the solution $yK_1(y)$ to be $1$ at the origin $y=0$.}.
Thus we must have $\beta(\lambda)\rightarrow 0$ in the large $\lambda$ limit.

We will now examine the  implications of the properties of  the solutions of the differential equation 
derived in (\ref{b1soln}), (\ref{b2sol}), (\ref{bnsol})  as well as  (\ref{b1soln2}), (\ref{b2soln2})
and (\ref{b0soln2})
  on the retarded Green's function $g^1(\omega)$. In terms of the 
variable $y$  the Green's function is given by 
\begin{equation}
\label{grelim}
 g^{1}(\lambda) = - \frac{r_+^2}{L^3}  \lim_{y\rightarrow 0} \frac{\lambda^2}{y} 
\frac{1}{\alpha(\lambda){\phi^1}_{(1)} (y) +  \beta(\lambda)  {\phi^1}_{(2)}( y) }
\frac{d}{dy} \left( {\alpha(\lambda) {\phi^1}_{(1)} (y) + \beta(\lambda)  {\phi^1}_{(2)}( y) }  \right).
\end{equation}
Taking the $y\rightarrow 0$ limit and using the behaviour of the functions
near the origin we obtain
\begin{equation}
 g^1(\lambda)   = - \frac{r_+^2}{L^3} \left(  \lim_{y\rightarrow 0} \frac{\lambda^2}{y} 
{a_0}^{(1)}(y)   + \frac{2}{3} (-2k_1+k_2+k_3)  + 2  \frac{ \beta(\lambda)}{ \alpha(\lambda)} \right).
\end{equation}
Note that the above expression is valid for all values of $\lambda$. 
Taking the $\lambda\rightarrow \infty$
limit we are left with 
\begin{equation} \label{limgl}
 \lim_{\lambda\rightarrow \infty}  
 g^1(\lambda)  =- \frac{r_+^2}{L^3} \left(  \lim_{y\rightarrow 0} 
 \frac{\lambda^2}{y} 
{a_0}^{(1)}(y)   + \frac{2}{3} (-2k_1+k_2+k_3)   \right) ,
\end{equation}
where we have used (\ref{basymch}).  The rest of the terms in  $g^1$ is subleading in the 
$\lambda\rightarrow \infty$ limit. 
 As we have mentioned earlier, the leading 
 contribution in the $\lambda$ expansion is identical to the zero temperature  case.  
 This has a $\log(r)$ divergence which has to be removed by the counter term. 
 Since this divergence is independent of the temperature we have the 
 relation
 \begin{equation}
 \label{count}
 G^1_{{\rm counter}} (\omega, T) = G^1_{{\rm counter}}( \omega, 0). 
 \end{equation}
 There is 
also a constant term due to the first order correction ${a_1}^{(1)}$. 
Thus the Green's function does not satisfy `property 2' and needs to be regulated. 

To regulate the Green's function we consider
\begin{equation}
 \delta G^{1}(\omega) = G^1(\omega, T) - G^1(\omega, 0)  - 
\frac{1}{e^2} \frac{2r_+^2}{3L^3} (-2k_1+k_2+k_3) .
\end{equation}
where  $e^2$ is defined in (\ref{imgreen}). 
Now using   (\ref{limgl})  as well as (\ref{count})  we have 
\begin{equation}
  \delta G^1(\omega) \rightarrow 0, \qquad \omega\rightarrow i \infty.
\end{equation}
We have essentially subtracted the divergent and the constant term in $G_R(\omega)$ so that 
`property 2' is true on $\delta G_R(\omega)$,  which can then be used to obtain 
the sum rule. 

\subsection{The sum rule}

From the above discussion we see that $\delta G_R(\omega)$ satisfies both
`property 1' and `property 2'.  Also since the subtracted constant in defining
$\delta G^1(\omega)$ is real we have the relation
\begin{equation}
{\rm Im } \delta G^1(\omega) =  \rho^1 (\omega)_T   -\rho^1(\omega)_{T=0}, 
\end{equation}
where
\begin{equation}
\rho^1(\omega)_T = {\rm  Im } G^1(\omega, T) , \qquad
\rho^1 (\omega)_{T=0} = {\rm Im} G^1(\omega, 0). 
\end{equation}
Thus using analyticity properties of $\delta G^1(\omega)$ we obtain the sum rule
\begin{equation}
 \int_{-\infty}^\infty \frac{d\omega}{\pi\omega}
\left( \rho^1(\omega) - \rho^1_{T=0} (\omega) \right)  = \delta G^1(0), 
\end{equation}
where  the RHS of the sum rule is given by 
\begin{equation}
\delta G^1(0) = G^1(0, T) - \frac{1}{e^2} \frac{2r_+^2}{3L^3} (-2k_1+k_2+k_3). 
\end{equation}
 The  zero frequency Green's function can  be related to 
the conductivity  as given in equation (\ref{solmatcon}). 
 From the analysis of the zero frequency limit  and 
the equation (\ref{imgreen}) we see that  this is given by the 
 residue of the pole of the imaginary part of the conductivity at zero frequency. 
 Therefore we have
 \begin{eqnarray}
  G_R^1(0, T) &=& \lim_{\omega\rightarrow 0} \omega {\rm Im} \,  \sigma^1(\omega)  , \\ \nonumber
  &=& \frac{1}{e^2}  \frac{r_+^2 k_1}{L^3}. 
  \end{eqnarray}
 Thus the sum rule  can be written as 
 \begin{equation}\label{ucsum1}
 \int_{-\infty}^\infty \frac{d\omega}{\pi\omega}
\left( \rho^1(\omega) - \rho^1_{T=0} (\omega) \right) 
=  \lim_{\omega\rightarrow 0} \omega {\rm Im} \,  \sigma^1(\omega)  
- \frac{1}{e^2} \frac{2r_+^2}{3L^3} (-2k_1+k_2+k_3)  . 
\end{equation}
Similarly  studying the Green's functions $G_R^2(\omega)$ and $G_R^3(\omega)$
we obtain the following sum rules for their spectral densities. 
\begin{eqnarray}
\label{ucsum23}
\int_{-\infty}^\infty  \frac{d\omega}{\pi\omega}
\left( \rho^2(\omega) - {\rho^2}_{T=0} (\omega) \right) = 
  \lim_{\omega\rightarrow 0} \omega {\rm Im} \,  \sigma^2 (\omega)  -
\frac{1}{e^2} \frac{2r_+^2}{3L^3} (k_1-2k_2+k_3), \nonumber\\
\int_{-\infty}^\infty  \frac{d\omega}{\pi\omega}
\left( \rho^3(\omega) - {\rho^3}_{T=0} (\omega) \right)  = 
  \lim_{\omega\rightarrow 0} \omega {\rm Im} \,  \sigma^3 (\omega)  -
 \frac{1}{e^2} \frac{2r_+^2}{3L^3} (k_1+k_2-2k_3). \nonumber \\
\end{eqnarray}
Note that at zero chemical  potentials all terms in 
the RHS of the the sum rules in (\ref{ucsum1}), (\ref{ucsum23}) 
vanish. Thus they reduce to the ones derived in \cite{Baier:2009zy} and \cite{Gulotta:2010cu}.

\section{Structure constants from sum rules}

From the derivation of the sum rule we see that the RHS of the sum rule in 
(\ref{ucsum1})  and (\ref{ucsum23}) 
consists of two terms. One term is the zero frequency dependence of the 
Green's function, this is determined by hydrodynamics. 
The zero frequency contribution is due to  the residue of the  pole in the imaginary part of the 
conductivity. 
The second term is from the finite term in the high frequency behaviour of the 
Green's function. 
As emphasized in \cite{David:2011hy} the  finite terms that arise in the sum 
rule  can be understood by examining the operator product expansion of the 
currents  involved in the  corresponding Green's function.  In this section 
we will examine the OPE  of the R-currents.  We will show that the 
finite term in the sum rule which arises from the high frequency behaviour
of the Green's function is due to the presence of chiral primary operators 
of dimension 2. These operators gain expectation values in the 
presence of chemical potential. From this  analysis we 
obtain the structure constants involving  two R-currents and the chiral primary. 
We then obtain these structure 
constants from  Witten diagrams using the 
methods developed in \cite{Freedman:1998tz} and show that they agree what that obtained 
from the sum rule.

\subsection{OPE and high frequency behaviour}

Here the OPE of interest is that of the R-currents which have conformal dimension
$3$ in four dimensions.  
The leading terms in this  OPE are given by 
\begin{equation}
\label{opejj}
 J_\mu^i(x)J_\nu^j(0) \sim \frac{ {\cal C}  \delta_{ij} I_{\mu\nu} (x)  }{x^6} 
+{\cal{A}}_{\mu\nu} C_{ij}^{\;\;\hat k} {\cal{O}}_{\hat k} (0) +
{\cal{B}}^{ij; \rho}_{\mu\nu; k} J^k_\rho(0) + \cdots, 
\end{equation}
where $\mu,\nu$ and $\eta$ are the  space time indices and $i, j $ and $k$ are the R-symmetry  indices.
The hatted indices $\hat k$ run over  the scalar operators in the theory. 
The tensor $I_{\mu\nu}(x)$ is given by 
\begin{equation} \label{defit}
 I_{\mu\nu}(x) = \delta_{\mu\nu}  - 2 \frac{x_\mu x\nu}{x^2}. 
\end{equation}
Note that for the purpose of analyzing the OPE  we work in   Euclidean signature.
The OPE given in (\ref{opejj}) can contain higher tensor operators for example,  the 
stress tensor.  
But as we will see,  these operators have higher dimensions and will not 
be relevant for our analysis.  
The tensor structure  ${\cal A}_{\mu\nu}$ can be obtained
by examining the three point function  $ \langle J_\mu^i J_\nu^j {\cal O}_{\hat k} \rangle $ and 
$C_{ij}^{\;\;\hat k} $ are the structure constants. 

Before we determine the 
tensor structure ${\cal A}_{\mu\nu}$, 
it  is easy to see that the ${\cal A}_{\mu\nu}$ must scale
as $x^{\Delta-6}$ where $\Delta$ is the conformal dimension of the scalar
operator ${\cal O}_{\hat k} $.  
Let us  now take the Fourier transform on both sides of the 
OPE in (\ref{opejj}) with spatial momentum 
$q=0$. Then by a simple scaling analysis as done in 
\cite{David:2011hy},  it is easy to see that the only operator which can contribute to 
the constant  term as the frequency $\omega$ in the Fourier transform is 
scaled to infinity  is $\Delta =2$. 
For $\Delta \neq 2$, the term scales as
 as $ \omega^{2- \Delta }$.  Thus these terms diverge  for $\Delta <2$ and are subleading 
when $\Delta >2$.    Note that the leading term 
in (\ref{opejj}) is proportional to the central
charge ${\cal C}$ scales like $x^{-6}$. This  diverges as $\omega^2$ 
on taking the Fourier transform. 
It is this divergence which needs to be subtracted to 
regulate the Green's function.  

Let us now examine the term which is proportional
to the R-current in the RHS of the OPE. The tensor
${\cal  B}$ contains various space-time structures which 
can be found in \cite{Osborn:1993cr} from the three point function  of $
\langle J_\mu^i  J_\nu^j  J_\rho^k \rangle $.  From conformal invariance we see that  
 ${\cal  B}$ scales as 
 $x^{-3}$. 
Using the same reasoning we see that the term proportional to the 
current on the RHS of the OPE is subleading as $\omega\rightarrow \infty$. 
 Similar reasoning allows us to conclude that 
tensor operators of higher dimension will not contribute in the large frequency limit. 

We are interested in  the expectation value of the Green's function in the thermal state
and at finite chemical potential. Thus taking expectation values on both sides of the 
OPE  in (\ref{opejj}),   we conclude that the term which is finite as $\omega\rightarrow\infty$ 
arises from the presence of expectation values of operators of dimension $2$ in the OPE. 
From the structure of the OPE  in (\ref{opejj}), 
we see that given the finite terms in the large $\omega$ expansion 
we can extract the structure constant $C_{ij}^{\;\;\hat k} $ if one has the knowledge of the 
expectation value of the scalar ${\cal O}_{\hat k}$. 
Our goal is to use the sum rules derived in (\ref{ucsum1}) and (\ref{ucsum23}) 
and extract out the structure constants $C_{ij}^{\;\;\hat k}$. For this purpose we we first will evaluate the 
expectation values of the operators of dimension 2 which are turned on in the  theory 
at finite chemical potential.  We  will then compare these structure constants evaluated  
using the conventional approach of \cite{Freedman:1998tz} and show that they  indeed do agree. 

To begin with, let us be more precise about  the structure of the tensor ${\cal A}_{\mu\nu}$. 
Consider the three point function of two currents with a scalar operator of dimension $2$. 
Conformal invariance constrains this to the form given by \cite{Anselmi:1996dd}
\begin{equation}
\langle J_\mu^i(x) J_\nu^j(y) {\cal{O}}_{\hat k} (z)\rangle = C_{ij \hat k}  \frac{4I_{\mu\nu}(x-y)-2 I_{\mu\rho}(x-z)I_{\rho\nu}(z-y)}{(x-y)^4(x-z)^4}. 
\end{equation}
The tensor $I_{\mu\nu}$ is defined in (\ref{defit}). 
The leading term in the short distance expansion when $x\rightarrow y$ of this 
three point function  is given by 
\begin{eqnarray}
\label{sdexp}
 \langle J_\mu^i(x)J_\nu^j(y){\cal{O}}_{\hat k}(z)\rangle 
&=&C_{ij\hat k} \frac{4I_{\mu\nu}(x-y)-2 I_{\mu\rho}(x-z)I_{\rho\nu}(z-y)}{(x-y)^4(x-z)^4},  \\  \nonumber
&=&  C_{ij\hat k}   \frac{ \hat{\cal{A}}_{\mu\nu}(s) }{(x-z)^4}  + \ldots,
\end{eqnarray}
where $s^\mu = (x - y)^\mu$ and 
\begin{eqnarray}
 \hat{\cal{A}}_{\mu\nu}(s) &=& -\frac{8 }{s^4}(\frac{s_\mu s_\nu}{s^2}-\frac{1}{4} \delta_{\mu\nu}),\nonumber\\
&=& -\partial_\mu\partial_\nu \frac{1}{s^2}. 
\end{eqnarray}
Note that the since $ \hat{\cal{A}}_{\mu\nu}(s)  = O(1/s^4)$ it is not  well defined 
as a distribution in 4 dimensions. Thus we need to regularize  this 
by introducing  a delta function following the methods of 
\cite{Osborn:1993cr,Freedman:1991tk,Freedman:1992tz,Latorre:1993xh}. 
Therefore we consider
\begin{eqnarray}
  {{\cal{A}}_{\mu\nu}}(s) = 
- \left( \partial_\mu\partial_\nu \frac{1}{s^2} + 4\pi^2 K \delta_{\mu\nu} \delta^4 (s)\right). 
\end{eqnarray}
This expression  is identical to $\hat {\cal A}_{\mu\nu}$ for $s\neq 0$. We now determine 
the constant $K$ using Ward identities. 
As a consequence of invariance under the $U(1)^3$  R-symmetry we obtain 
  the Ward identity \cite{Osborn:1993cr} \footnote{Here we ignore the $U(1)^3$ anomaly since 
all our computations will be at zero momentum. }
\begin{equation}
\label{ward}
\partial^\mu   \langle J_\mu^i \rangle  =0
\end{equation}
Here $A_\mu^i $ are the sources corresponding to the currents $J_\mu^i$.  
Differentiating this identity with respect to $A_\nu^j (y) $ and  the source corresponding to the 
operator ${\cal O}_{\hat k}  (z) $ we obtain the Ward identity
\begin{equation}
 \partial^\mu \langle J _\mu^i (x) J_\nu^j (y) {\cal O}_{\hat k} (z) \rangle  =0. 
\end{equation}
Note that the operator ${\cal O}_{\hat k}$ is uncharged under the R-symmetry. 
Thus we must have 
\begin{eqnarray}
 \partial^{\mu} {{\cal{A}}_{\mu\nu}}(s) &=&-(K-1) 4\pi^2\partial_\nu \delta^4 (s)=0. 
\end{eqnarray}
This implies $K=1$ and 
the regularized expression  for  ${\hat{\cal{A}}_{\mu\nu}}(s)$ is  given by 
\begin{equation}\label{fintens}
   {{\cal{A}}_{\mu\nu}}(s) = -
 \left(  \partial_\mu\partial_\nu \frac{1}{s^2} + 4\pi^2 \delta_{\mu\nu} \delta^4 (s)\right).  
\end{equation}
Thus  contribution of the scalar operator of  dimension 2 in  that the OPE of the R-currents is  given by
\begin{equation}\label{finope}
 J_\mu^i (x) J_\nu^j(y)  \sim  C_{ij}^{\;\;\hat k}  {\cal A}_{\mu\nu} (s)  {\cal O}_{\hat k} (x) + \cdots. 
\end{equation}
Note that now $\hat k$ runs over all operators of dimension $2$ in the theory. 
Inserting the operator ${\cal O}_{\hat l} (z)$ on both sides of the above equation and comparing 
it with the three point function in (\ref{sdexp}), we obtain 
\begin{equation} \label{loing}
 C_{ij\hat l}  = C_{ij}^{\;\;\hat k} g_{\hat k \hat l},  
\end{equation}
where the constants $ g_{\hat k \hat l }$ are defined by the two point function 
\begin{equation}
 \langle {\cal O}_{\hat k} (x) {\cal O}_{\hat l} (z) \rangle  = \frac{g_{\hat k \hat l}} { ( x - z)^4 }.  
\end{equation}

Let us now examine the contribution of all  the operators of dimension $2$  that 
appear in the OPE to the finite
terms  in the $\omega\rightarrow \infty$ of the  Euclidean Green's function. 
For this we evaluate the following
\begin{eqnarray} \label{opeuc}
\lim _{\omega\rightarrow \infty}  \delta G^{i}_E ( \omega) &=& \int d^4 x e^{-i\omega t } \left( 
\langle J_x^i (x) J_x^i ( 0) \rangle_{T\neq 0} -  \langle J_x^i (x) J_x^i ( 0) \rangle_{T= 0} \right), 
\\ \nonumber
&=& C_{ii}^{\;\;\hat k} \langle {\cal O}_{\hat k}  (0)  \rangle_{T\neq 0}  \int d^3 x e^{-i\omega t } 
{\cal A}_{xx}( x) , \\ \nonumber
&=& -4\pi^2 C_{ii}^{\;\;\hat k} \langle {\cal O}_{\hat k}  (0)  \rangle_{T}. 
\end{eqnarray}
To obtain the second line we have substituted the OPE derived in (\ref{finope}) 
and the form for ${\cal A}$ given in (\ref{fintens}). We have also  
used the fact that expectation value of the identity operator is normalized to 
$1$ in both at zero and at finite temperature, this leads to the cancellation 
of the $\omega^2$ divergence from the term proportional to the central charge 
${\cal C}$ in the OPE. 
Note that the equation in (\ref{opeuc})  is a relation for the Euclidean propagator.
The Euclidean propagator can be  analytically continued to the retarded 
Minkowski propagator in the whole of the upper half $\omega$ plane \cite{Son:2002sd} using the 
relation
\begin{equation}
G(i\omega, q) = - G_E( \omega, q)
\end{equation}
This leads to the following behaviour for the retarded two point function of interest in 
the limit $\omega\rightarrow i \infty $
\begin{equation}
\label{finter1}
 \lim_{\omega\rightarrow i\infty}  G^{i} _T (\omega) - G^{i}_{T=0} (\omega) 
=  4\pi^2 C_{ii}^{\;\;\hat k} \langle{\cal  O}_{\hat k}  (0)  \rangle_{T}. 
\end{equation}
It is important to point out the analysis in this subsection has been performed 
entirely based on the conformal invariance of the dual theory. It does not rely 
on holography. 

\subsection{Reading  the structure constants from the sum rule}

From the analysis of the behaviour of the OPE we have concluded that the 
finite terms  in the sum rule from  $\omega\rightarrow i\infty$  limit arises
because of certain operators of dimension $2$  in the OPE. 
From the equation in (\ref{finter1}),  we see that once  we know the finite term
in the Greens' function  in the $\omega \rightarrow i\infty$, 
and the expectation values of the operators  of dimension 
$2$ in the thermal state,  we can read out the structure constants involved. 
We will now perform this analysis from the sum rules in (\ref{ucsum1}) and  (\ref{ucsum23})
derived holographically. 

The operators of dimension $2$ which are relevant for our analysis are dual to the 
$2$ independent scalars present in the background given in 
 (\ref{scalval}). 
We parametrize  $X^i$'s in terms of these independent scalars by   
\begin{eqnarray}
\label{xphi1c}
& & X^1 = \exp\left(  -\frac{1}{\sqrt{6}} \vartheta_1- \frac{1}{\sqrt{2}}  \vartheta_2 \right) ,\qquad 
X^2 = \exp\left(   -\frac{1}{\sqrt{6}}\vartheta_1+ \frac{1}{\sqrt{2}}  \vartheta_2    \right), \\ \nonumber 
& &  X^3 = \exp\left( \frac{2}{\sqrt{6}} \vartheta_1 \right) .
\end{eqnarray}
Substituting these  field redefinitions in the action (\ref{cgaction}) and 
expanding the scalar potential 
\begin{equation}
 V =\frac{4}{L^2}\sum_{i=1}^{3}\frac{1}{X_i}, 
\end{equation}
to quadratic order in $\vartheta_i$ it can be shown that the masses of both is given by 
\begin{equation}
 m^2 L^2 =-4.
\end{equation}
Using the  mass-dimension relation for scalars
\begin{equation}
 \Delta( \Delta -4) = m^2L^2, 
\end{equation}
we see that the fields $ \vartheta_i$ corresponds to operators of dimension $2$ in the field theory and 
saturate the Breitenlohner-Freedman bound.  
From  the  ${\cal N}=4$ field theory  point of view 
 these operators correspond to  two linear combinations  of the 
following chiral primaries
\begin{equation} 
{\rm Tr} ( X\bar X), \qquad  { \rm Tr} ( Y\bar Y), \qquad  {\rm Tr}(Z\bar Z) , 
\end{equation}
where $X, Y, Z$ are 3 complex fields constructed out of the $6$ scalars in ${\cal N}=4$ Yang-Mills. 
These are the three scalars which are un-charged under the Cartans of $SO(6)$. 
Note that the combination $ {\rm Tr} ( X\bar X) +   { \rm Tr} ( Y\bar Y) +   {\rm Tr}(Z\bar Z) $
is the Konishi scalar and therefore not a  chiral primary and is to be excluded.  
We will denote the two chiral primaries dual to the field $\vartheta_1, \vartheta_2$ as 
${\cal  O}_1$ and ${ \cal O}_2$ respectively. 

it is easy to read out the expectation values of the 
operators ${\cal O}_1$ and ${\cal  O}_2$  at finite chemical potential from the 
background given in (\ref{chgmet}) and (\ref{scalval}) 
following the procedure detailed in \cite{Marolf:2006nd}.
See \cite{Witten:2001ua, Berkooz:2002ug} for earlier work. Details of this has been 
repeated in \cite{David:2011hy}. Here we write down the results for the expectation values. 
\begin{eqnarray}
\label{expect}
  \langle {\cal O}_1\rangle &=& 
\frac{N^2}{8\pi^2 } \frac{r_+^2}{L^4}\frac{2}{\sqrt{6}} ( k_1+k_2-2k_3), \\ \nonumber
\langle {\cal O}_2 \rangle &=&  
\frac{N^2} {8\pi^2 }\frac{r_+^2}{L^4} \frac{2}{\sqrt{2}} ( k_1-k_2).
\end{eqnarray}
Note that these expectation values have mass dimension 2 since the operators 
${\cal O}_1,  {\cal O}_2$ have
mass dimension 2 and  they are  constant in space time.
We have  normalized   the operators following \cite{Klebanov:1999tb}, that is by dividing scalar operators
by $\Delta - d/2$ where $\Delta$ is the conformal dimension of the operator
and $d$ is the space time dimension \footnote{See discussion below equation (2.21)
of \cite{Klebanov:1999tb}.}. This explains
the additional factor of $2$ in (\ref{expect}) when compared to equation (4.70) of 
\cite{David:2011hy}.  
We can now compare the large frequency terms in the sum rule  
 (\ref{ucsum1}), (\ref{ucsum23})  and the expectation values of the scalars in 
(\ref{expect})  to  rewrite the RHS of the sum rules   as  
\bea
\label{rggfo}
\delta G^{1} (0) &=&  \lim_{\omega\rightarrow 0} \omega {\rm Im} \,  \sigma^1(\omega) 
+\f{1}{\sqrt{6}L^2}\langle{\cal{O}}_1\rangle +\f{1}{\sqrt{2}L^2} \langle{\cal{O}}_2\rangle,\nonumber\\
\delta G^2 (0) &=& \lim_{\omega\rightarrow 0} \omega {\rm Im} \,  \sigma^2(\omega) 
+\f{1}{\sqrt{6}L^2}\langle{\cal{O}}_1\rangle -\f{1}{\sqrt{2}L^2} \langle{\cal{O}}_2\rangle,\nonumber\\
\delta G^3 (0) &=& \lim_{\omega\rightarrow 0} \omega {\rm Im} \,  \sigma^3(\omega)  -
\f{2}{\sqrt{6}L^2} \langle{\cal{O}}_1\rangle.\nonumber\\
\eea
where  we have used  the expression for $e^2$ given in (\ref{imgreen}). 
Comparing the contribution from the 
higher frequency part in (\ref{rggfo}) and (\ref{finter1}) we  find the  following values for   the 
structure constants \footnote{The finite terms in the Green's function at  the high frequency 
 is negative of that in the sum rule (\ref{rggfo}) since this term is subtracted in the 
renormalized Green's function. }
\bea
\label{zeta}
C_{11}^{\;\;\hat 1} &=&-\f{1}{(2\pi)^2 L^2\sqrt{6}},\qquad 
C_{11}^{\;\;\hat 2} = -\f{1}{(2\pi)^2L^2\sqrt{2}},\nonumber\\
C_{22}^{\;\;\hat 1} &=&-\f{1}{(2\pi)^2L^2\sqrt{6}},\qquad 
C_{22}^{\;\;\hat 2} =\f{1}{(2\pi)^2 L^2\sqrt{2}},\nonumber\\
C_{33}^{\;\;\hat 1} &=&\f{2}{(2\pi)^2L^2\sqrt{6}},\qquad
C_{33}^{\;\;\hat 2} =0.
\eea
Note that here the first two indices in the subsripts of $C$ are the R-symmetry 
indices, while the third hatted index  labels the operators of dimension $2$.

\subsection{Structure constants from  Witten diagrams}

In this section we use the the standard AdS/CFT prescription put forward in 
 \cite{Witten:1998qj}  to evaluate the 
the three point function $\langle J_\mu^i J_\nu^j {\cal O}_{\hat k} \rangle$ 
This will enable to us to obtain the structure constants $C_{ij\hat k}$. 
We can then compare them to that obtained from the sum rule listed in (\ref{zeta}). 
To obtain the structure constants using the conventional 
AdS/CFT prescription we follow the methods developed in 
\cite{Freedman:1992tz}. From the action given in (\ref{cgaction}) we see that the 
 three point function 
of two R-currents with the operators of dimension 2 
 can be evaluated by examining the 
cubic interaction of the scalars  with the gauge fields. 
This interaction occurs in the gauge kinetic term. Rewriting the scalars $X^i$ in terms of the 
the two independent scalars $\vartheta_1, \vartheta_2$ using the equations in 
(\ref{xphi1c}) we obtain the following interaction
\begin{equation}
\label{vvphi}
 \frac{1}{4 e^2} \vec a^i \cdot
\int d^5 x \sqrt{g}  g^{\mu\rho} g^{\nu\sigma}  \vec \vartheta  F_{\mu\nu}^{i} F_{\nu\rho}^i, 
\end{equation}
where  we have  organized the two scalars $\vartheta_1, \vartheta_2$ 
into a two dimensional vector and 
the  two dimensional vectors  $\vec a^i $  with $i =1, 2, 3 $ are  given by 
\begin{eqnarray}\label{d3av}
\vec a^1 = 2 \left( \frac{1}{\sqrt{6}}, \frac{1}{\sqrt{2} } \right) , \quad 
\vec a^2 = 2 \left( \frac{1}{\sqrt{6} }, - \frac{1}{\sqrt{2} } \right), \quad
\vec a^3 = 2 \left( - \frac{2}{ \sqrt{6} }, 0  \right) . 
\end{eqnarray}
In (\ref{vvphi}) summation over $i$ is implied. 
The three point function of two currents with an operator of dimension $\Delta$ in a
 $d$ dimensional conformal field theory  
is obtained holographically using the methods of \cite{Freedman:1992tz} in  appendix A. 
Applying the equation (\ref{struc}) to the cubic interaction given in (\ref{vvphi}) together
with $d=4$  and $\Delta =2$ 
we obtain  the following values for the structure constants. 
\begin{equation}\label{lower}
C_{ii\hat j} = - a^{i }_{\hat j} {\cal K}_ {4}(2),  
\end{equation}
where ${\cal K}$ is defined in (\ref{struc}) and the subscript with the hatted index 
refers to the component of the vector $\vec a^i$ defined in (\ref{d3av}). 
The structure constants evaluated from the sum rules given in 
(\ref{zeta})  have one one raised index which are related to the ones in 
(\ref{lower})  by contraction with the constants that appear in the 
two point function of the scalars as given in (\ref{loing}). Given the action for the 
scalars in (\ref{cgaction}),  
the two point function of the corresponding  with dimension $\Delta$ have been 
evaluated holographically in (\ref{twopt}). 
This is given by 
\begin{equation}\label{zmet}
 g_{\hat i\hat j} = \delta_{\hat i\hat j} { \cal G}_{4} (2). 
\end{equation}
The Kronecker delta results from the fact that the scalars $\vec\vartheta$ have a
canonically normalized Kinetic term.
Therefore using (\ref{loing})  together with (\ref{lower}) and (\ref{zmet}) we obtain
\begin{equation}
C_{ii}^{\;\;\hat j} = - a^i_{\hat j} \frac{ { \cal K}_{4} (2) } { { \cal G}_{4} (2) }. 
\end{equation}
Substituting the values of the normalizations and the components of the 
vector $\vec a^i$ we obtain
\begin{eqnarray}
\label{zetaw}
C_{11}^{\;\;\hat 1} &=&  -\f{1}{(2\pi)^2 L^2\sqrt{6}},\qquad 
C_{11}^{\;\;\hat 2} =  - \f{1}{(2\pi)^2L^2\sqrt{2}},\nonumber\\
C_{22}^{\;\;\hat 1} &=& - \f{1}{(2\pi)^2L^2\sqrt{6}},\qquad 
C_{22}^{\;\;\hat 2} =  \f{1}{(2\pi)^2 L^2\sqrt{2}},\nonumber\\
C_{33}^{\;\;\hat 1} &=&\f{2}{(2\pi)^2L^2\sqrt{6}},\qquad
C_{33}^{\;\;\hat 2} =0.
\end{eqnarray}
Comparing (\ref{zeta}) and (\ref{zetaw}) we see that the structure constants derived from the 
sum rule precisely coincides with that evaluated using the Witten diagrams. 

\section{R-charge sum rules for M2 and M5-branes}

In this section we will holographically derive the R-charge sum rules for the case of M2-branes and 
M5-branes.  We will not go through all the details as was done for the D3-brane,  but the 
same analysis can be repeated for these situations. 
We will only highlight the important  points in the derivation of the sum rules  so that 
the results can be presented.  
For the case of the M2 and the M5-branes there are finite terms in the sum rules which 
arises due to the high frequency behaviour of the relevant Green's function. 
We show that 
these finite terms are due to expectation values of operators  of dimension 1 and dimension 4 
respectively. 
We   then compute the structure constants of the R-currents with these operators form the 
finite terms in the sum rule and show that they indeed agree with that evaluated  from the corresponding
Witten diagrams. 

\subsection{M2-branes}

The theory on the M2-brane has an $SO(8)$ R-symmetry. Thus there are at the most $4$ chemical potentials
corresponding to the  Cartans of $SO(8)$ which can be turned on. 
The gravity dual of this solution is given in (\ref{ch4met}) and (\ref{valscalm2}). 
As for the case of D3-brane, the object of interest is the R-current retarded Green's function. 
To obtain this Green's function holographically we need to study the fluctuations  of the 
$4$ gauge fields dual to the R-currents corresponding to the Cartan of $SO(8)$. 
These equations are given by 
\begin{eqnarray}
\label{m2zeqn}
 \phi^{i\prime\prime} +\left(\f{f'}{f}+\f{2H'_i}{H_i}-\f{{\cal H}'}{{\cal H}} \right) \phi^{i\prime}
+ \frac{ L^4\omega^2}{r_+^2 f} {\cal H} \phi^i 
-(1+k_i)\m_i \sum_{j=1}^{4}\f{L^2}{r_+^2} \frac{u^2}{H_i^2 f}  \m_{j}(1+k_j) \phi^j  =0. \nonumber
\\
\end{eqnarray}
Here the prime is the derivative with respect to 
\begin{equation}
 u = \frac{r_+}{r}, 
\end{equation}
the indices $i$ now run from $1, \cdots 4$. The functions $f, H_i, \mu_i, {\cal H}$ are defined in 
(\ref{ch4met}).  $\phi^i(u)$ is the fluctuation of the gauge field in the $x$ direction for the 
Fourier mode with frequency $\omega$. We have set the spatial momentum of the fluctuations
to zero. 
To obtain the equations for the fluctuations we have followed the same  procedure
as that in the case of the D3-brane.  One can consistently turn on fluctuations of the 
gauge field in the $x$ direction and the metric component  in the $xt$ direction. Then 
we obtain a constraint equation if one examines the fluctuation in the $xu$ component 
of the Einstein equation.  This can be used to eliminate the fluctuation in the 
metric and obtain the set of equations given in (\ref{m2zeqn}). 

The important term in the sum rule for our analysis is the finite term in the
$\omega\rightarrow i\infty$ limit. This can be obtained by the same analysis as that
developed for the D3-brane. 
We first rescale the coordinates as 
\begin{equation}
 y  = \lambda u , \qquad i\lambda = \frac{L^2}{r_+} \omega . 
\end{equation}
Then the set of coupled equations in (\ref{m2zeqn}) becomes 
\begin{eqnarray}
\label{m2zeqn2}
\phi^{i\prime\prime} +\left(\f{f'}{f}+\f{2H'_i}{H_i}-\f{{\cal H}'}{{\cal H}} \right)\phi^{i\prime}
- \frac{{\cal H}}{f^2 } \phi^i
-(1+k_i) \m_i\sum_{j=1}^{4} \f{L^2}{r_+^2}\frac{y^2}{\l^4 H_i^2 f}  \m_{j}(1+k_j) \phi^j 
=0. \nonumber \\
\end{eqnarray}
where, now the derivatives are with respect to the variable $y$. 
Note that the equations decouple in the strict $\lambda\rightarrow\infty$ limit and 
they reduce to that of  gauge fields in pure $AdS_4$. 
Thus the leading term in the large $\lambda$ limit 
is again identical to the zero temperature situation.
The leading independent solutions are 
\begin{equation}
 \phi^i \sim e^{\pm y}
\end{equation}
Demanding that  the solution is well behaved at the origin of $AdS$ picks out 
 $e^{-y}$ as the zero temperature solution.
To obtain the asymptotic behaviour of the Green's function we define the variables
\begin{equation}
a(y) = \frac{\phi^{1\prime}(y)}{\phi^1(y)}, \quad
b(y) = \frac{\phi^{2\prime}(y)}{\phi^2(y)}, \quad
c(y) = \frac{\phi^{3\prime}(y)}{\phi^3(y)}, \quad
d(y) = \frac{\phi^{4\prime}(y)}{\phi^4(y)}. 
\end{equation}
We can then set up a perturbative expansion of each of the above functions 
as series in $1/\lambda$. 
We will discuss the leading equations for the function $a(y)$. 
We first expand $a(y)$ as a series in $1/\lambda$ as 
\begin{equation}
a(y) = a_0 +  \frac{a_1}{\lambda} + \frac{a_2}{\lambda^2}+ \cdots. 
\end{equation}
Substituting this expansion for $a$ in the non-linear equations determined from
 (\ref{m2zeqn2}) and matching powers of $1/\lambda$ we obtain the 
 the following equations for the leading coefficients.
 \begin{eqnarray}
 \label{lamexm2}
a_0' + a_0^2 -1  &=& 0, \\ \nonumber
a_1'+2a_0a_1+2k_1 a_0+(k_1+k_2+k_3+k_4)y &=&0
\end{eqnarray}
The two independent solutions for the first equation in (\ref{lamexm2}) are 
\begin{equation}
a_0^{(1)} = -1, \qquad a_0^{(2)} = 1. 
\end{equation}
The second solution corresponds to the growing solution $\phi^i \sim e^y$. But as we have 
mentioned above,  to have a well behaved solution at the origin in the zero temperature 
limit we must chose
the first solution $a_0^{(1)} =-1$. Evaluating the leading correction to $a_0^{(1)}$
we obtain
\begin{equation}
a_1^{(1)} = \f{1}{4}(-3k_1+k_2+k_3+k_4)+\f{y}{2}(k_1+k_2+k_3+k_4).
\end{equation}
The retarded Green's function is given by 
\begin{equation}
G^i (\omega, T) = - \frac{1}{e^2} \lim_{r\rightarrow\infty} \left.   \frac{r^2 \phi^{i\prime}}{L^2 \phi^i} 
\right|_{\phi^j_\infty=0, j\neq i} + G_{\rm{counter}}( \omega, T ) , 
\end{equation}
with
\begin{equation}
\frac{1}{e^2} = \frac{N^{\frac{3}{2}} \sqrt{2}}{24\pi L^2}. 
\end{equation}
Substituting the expressions for fluctuations $\phi^1$ in the $\omega\rightarrow i\infty$ limit
we obtain
\begin{equation}
G^1(\omega, T)|_{\omega\rightarrow i\infty}  =
 \f{1}{e^2} \lim_{y\rightarrow 0}
 \frac{r_+ }{L^2} \left( \lambda a_0^{(1)}(y)  + a_1^{(1)}(y)  \right) +G_{\rm{counter}} (\omega, T). 
 \end{equation}
 From our discussion we see that the divergent term in the frequency is 
 identical to the zero temperature limit of the the Green's function. 
 This motivates us to define the regularized Green's function as 
 \begin{equation} 
 \delta G^1(\omega) = G^1(\omega, T) - G^1(\omega, 0)  - \frac{r_+ }{4e^2 L^2}
 (-3k_1+k_2+k_3+k_4).
\end{equation}
With this regularization, the Green's function satisfies the properties necessary to 
obtain the sum rule. Thus the sum rule for this Green's function is given by 
\begin{equation}\label{summ21}
 \int_{-\infty}^\infty  \frac{d\omega}{\pi\omega}
\left(\rho^1(\omega) - \rho^1_{T=0} (\omega) \right) =
\lim_{\omega\rightarrow 0} \omega {\rm Im}  \, \sigma^1(\omega) 
 - \frac{r_+ }{4e^2 L^2}(-3k_1+k_2+k_3+k_4).
 \end{equation}
 where $\rho^i  ={\rm Im}\, G^i$ and $\sigma^i$ is the corresponding conductivity. 
 Similarly, 
 the sum rules for the other diagonal components of the Green's function 
 are given by 
 \begin{eqnarray}\label{summ22}
 \int_{-\infty}^\infty  \frac{d\omega}{\pi \omega}
\left(\rho^2(\omega) - \rho^2_{T=0} (\omega) \right) =
\lim_{\omega\rightarrow 0} \omega {\rm Im}  \, \sigma^2(\omega) 
 - \frac{r_+ }{4e^2 L^2}(k_1 -3k_2+k_3+k_4), \nonumber \\ \nonumber
  \int_{-\infty}^\infty  \frac{d\omega}{\pi\omega}
\left(\rho^3(\omega) - \rho^3_{T=0} (\omega) \right) =
\lim_{\omega\rightarrow 0} \omega {\rm Im}  \, \sigma^3(\omega) 
 - \frac{r_+ }{4e^2 L^2}(k_1 + k_2- 3k_3+k_4), \\ \nonumber
 \int_{-\infty}^\infty  \frac{d\omega}{\pi\omega}
\left(\rho^4(\omega) - \rho^4_{T=0} (\omega) \right) =
\lim_{\omega\rightarrow 0} \omega {\rm Im}  \, \sigma^4(\omega) 
 - \frac{r_+ }{4e^2 L^2}(k_1 +k_2+k_3 -3k_4). \\
 \end{eqnarray}
 
 As discussed in the case of the D3-brane  
 let us now examine the terms in the OPE of the R-currents to understand the 
 high frequency terms in the RHS of the sum rule.  The OPE of two R-currents 
 in a 3 dimensional CFT is given by 
 \begin{equation}
\label{opejj2}
 J_\mu^i(x)J_\nu^j(0) \sim \frac{ {\cal C}  \delta_{ij} I_{\mu\nu} (x)  }{x^4} 
+{\cal{A}}_{\mu\nu} C_{ij}^{\;\;\hat k} {\cal{O}}_{\hat k} (0) +
{\cal{B}}^{ij; \rho}_{\mu\nu; k} J^k_\rho(0) + \cdots, 
\end{equation}
The R-currents have dimension $2$ and therefore from a similar  scaling analysis discussed
for the case of the D3-branes we see that the finite terms in the high frequency limit
arise from scalar operators of dimension $1$ in the OPE. 
For operators of dimension 1, the regularized tensor  ${\cal A}$ is given by 
\begin{equation}
{\cal A}_{\mu\nu} (s) = -2 \left( \partial_{\mu}\partial_\nu \frac{1}{s} + 
4\pi \delta_{\mu\nu} \delta^3 (s) \right). 
\end{equation}
Taking the Fourier transform of the OPE and going over to Minkowski space we 
obtain the following relation for the high frequency behaviour of the retarded 
Green's function
\begin{equation} \label{him2}
\lim_{\omega\rightarrow\infty} ( G^i(\omega, T) - G^i(\omega, 0)) =
8\pi C_{ii}^{\; \hat k} \langle {\cal O}_{\hat k} (0) \rangle_T, 
\end{equation}
where ${\cal O}_{\hat k}$ are the operators of dimension 1 in the theory. 

For the M2-brane theory,  there are $3$ operators of dimension $1$ corresponding to the 
scalars $\vartheta^{\hat i}$ given in  (\ref{indpsc4}). Note that in this case the scalars
obey alternate quantization. 
  The expectation values of these operators in the 
charged M2-brane background is given by 
 \begin{eqnarray}
 \langle {\cal O}_1 \rangle
 &=& \left( \frac{N^{\frac{3}{2}}\sqrt{2}}{24\pi}\right)
  \frac{r_+}{ L^2}(k_1+k_2-k_3-k_4),\nonumber\\
\langle {\cal O}_2 \rangle
&=& \left( \frac{N^{\frac{3}{2}}\sqrt{2}}{24\pi }\right)
\frac{r_+}{ L^2}(k_1-k_2+k_3-k_4),\nonumber\\
 \langle {\cal O}_3 \rangle 
 &=& \left( \frac{N^{\frac{3}{2}}\sqrt{2}}{24\pi}\right)
\frac{r_+}{L^2}(k_1-k_2-k_3+k_4).
\end{eqnarray}
We can now rewrite the high frequency terms in the RHS of the sum rules given in 
(\ref{summ21}) and (\ref{summ21}) as 
\begin{eqnarray}
\label{srm2og}
 \delta G^1(0, T) &=&  \lim_{\omega\rightarrow 0} \omega {\rm Im}  \, \sigma^1(\omega) +
\frac{1 }{4 L^2}\left( \langle{\cal O}_1\rangle+\langle{\cal O}_2\rangle+\langle{\cal O}_3\rangle\right),\nonumber\\
 \delta G^2(0, T) &=& \lim_{\omega\rightarrow 0} \omega {\rm Im}  \, \sigma^2(\omega) 
+\frac{1 }{4 L^2}\left( \langle{\cal O}_1\rangle-\langle{\cal O}_2\rangle-\langle{\cal O}_3\rangle\right) ,\nonumber\\
 \delta G^3(0, T) &=&\lim_{\omega\rightarrow 0} \omega {\rm Im}  \, \sigma^3(\omega) 
+\frac{1 }{4 L^2}\left(- \langle{\cal O}_1\rangle+\langle{\cal O}_2\rangle-\langle{\cal O}_3\rangle\right) ,\nn\\
 \delta G^4(0, T) &=&\lim_{\omega\rightarrow 0} \omega {\rm Im}  \, \sigma^4(\omega) 
+\frac{1 }{4 L^2}\left(- \langle{\cal O}_1\rangle-\langle{\cal O}_2\rangle+\langle{\cal O}_3\rangle\right).
\end{eqnarray}
Comparing the equation (\ref{him2}) with the high energy contribution in the sum rules in 
(\ref{srm2og}) 
we read out the following values for the structure
constants
\begin{eqnarray}
\label{zsm2}
C_{11}^{\;\hat 1} &=&\f{-1}{32\pi L^2},\qquad 
C_{11}^{\; \hat 2} =\f{-1}{32\pi L^2},\qquad
C_{11}^{\; \hat 3} =\f{-1}{32\pi L^2},\nonumber\\
C_{22}^{\;\hat 1}&=&\f{-1}{32\pi L^2},\qquad
 C_{22}^{\;\hat 2} =\f{1}{32\pi L^2},\qquad
C_{22}^{\;\hat 3}  =\f{1}{32\pi L^2},\nonumber\\
C_{33}^{\;\hat 1}&=&\f{1}{32\pi L^2},\qquad 
C_{33}^{\;\hat 2}=\f{-1}{32\pi L^2},\qquad
C_{33}^{\;\hat 3}  =\f{1}{32\pi L^2},\nonumber\\
C_{44}^{\;\hat 1} &=&\f{1}{32\pi L^2},\qquad 
C_{44}^{\;\hat 2} =\f{1}{32 \pi L^2},\qquad
C_{44}^{\;\hat 3} =\f{-1}{32\pi L^2},\nonumber\\
\end{eqnarray}

We will now compare these values for the structure constants to that evaluated using the 
Witten's prescription.  Following the same procedure as in the case of the D3-brane
we obtain 
\begin{equation}\label{wittm2}
C_{ii}^{\hat j} = - a_{\hat j}^{i} \frac{{\cal K}_3 (1) }{ {\cal G}_3 (1)}
\end{equation}
where the  $a_{\hat j}^{i}$ refer to the $\hat j$th component of the 
vector $ \vec a^i$ given in (\ref{indpsc4}). 
The normalization constants in (\ref{wittm2}) are read out from 
(\ref{struc}) and (\ref{twopt}).  
Substituting these constants, we  indeed obtain  the structure constants
given in (\ref{zsm2}).

\subsection{M5-brane}

The theory on the M7-brane has a $SO(5)$ R-symmetry. 
Therefore there are at most $2$ chemical potentials corresponding to the 
2 Cartans that can be  turned on. The gravity background dual to this 
system is given in (\ref{ads7}) and (\ref{scalm5}). 
To obtain the R-current retarded correlators we study the 
fluctuations of the two  gauge fields in this background. 
They obey the following equations
\begin{equation}
\label{m5zeqn}
\phi^{i\prime\prime}
+\left(\log  \frac{H_i^2f}{z^3 {\cal H }}  \right)'\phi^{i\prime}
+ \frac{L^4 \omega^2}{r_+^2} {\cal H}{f^2} \phi^i 
-\f{16\mu_i z^8(1+k_i)}{H_i^2 f} \sum_{j=1}^{2} \mu_{j}(1+k_j) \phi^j  =0,
\end{equation}
where the prime is with respect to $z$ defined as
\begin{equation}
 z = \frac{r_+}{r}. 
\end{equation}
The functions $f, {\cal H}, H_i, \mu_i$ are defined in (\ref{ads7}) and (\ref{chemm5}). 
Here just as in the case of  the D3-brane and the M2-brane,
$\phi^i$  are the fluctuations of the gauge field in the $x$ direction for the 
Fourier mode with frequency $\omega$. 
We have set all fluctuations except that of the metric in the $xt$ direction to zero. 
This metric fluctuation is eliminated using the constraints  to obtain 
the equations in  (\ref{m5zeqn}). 

To study the $\omega\rightarrow i\infty$ limit we rescale the coordinates as
\begin{equation}
 y = \lambda z, \qquad  i \lambda = \frac{L^2}{r_+} \omega. 
\end{equation}
Then the equations in (\ref{m5zeqn}) become
\begin{equation}
\label{m5zeqn2}
\phi^{i\prime\prime}
+\left(\log  \f{H_i^2f}{y^3 {\cal{H}}} \right)'\phi^{i\prime}
- \lambda^2 {\cal H}{f^2} \phi^i 
-\f{16\mu_i y^8(1+k_i)}{\lambda^{10} H_i^2 f} \sum_{j=1}^{2} \mu_{j}(1+k_j) \phi^j  =0,
\end{equation}
where now the derivatives are with respect to $y$. 
Here again the equations decouple in the limit $\lambda\rightarrow\infty$
and reduce to the zero temperature pure $AdS_7$ situation.  
To obtain the large frequency behaviour of the Green's function we define
\begin{equation}
 a = \frac{\phi^{1\prime}(y)}{\phi^1(y)},  \qquad  b = \frac{\phi^{2\prime}(y)}{\phi^2(y)}. 
\end{equation}
One can then set up a perturbative expansion for each of these 
functions in  powers of $1/\lambda$.  We will briefly discuss the 
leading equations for the function $a(y)$. We expand this function as
\begin{equation}
 a(y) = a_0 + \frac{1}{\lambda^4} a_1 + \frac{1}{\lambda^6} a_2 + \cdots. 
\end{equation}
Substituting this expansion in the non-linear equation for $a$ determined from 
(\ref{m5zeqn2}), we obtain the following equations for the leading orders
\begin{eqnarray}
\label{expm5}
& & a_0'+a_0^2-\f{3}{y}a_0-1=0,\nn\\
& & a_1'+2a_0a_1-\f{3}{y}a_1+8 y^3k_1a_0+y^4(k_1+k_2)=0.
\end{eqnarray}
The two independent solutions for the first equation in \ref{expm5} are 
\begin{equation}
a_0^{(1)}=-\f{K_1(y)}{K_2(y)}, \qquad a_0^{(2)}=\f{I_1(y)}{I_2(y)}.
\end{equation}
Demanding that the 
solution is well behaved at the origin, $y\rightarrow\infty$,  at the leading
order in $\lambda$  picks out the first solution $a_0^{(1)}$. 
The leading correction to $a_0^{(1)}$.   is given by 
\begin{equation}\label{1stcorm5}
a_1^{(1)}=-4k_1y^3-(k_1+k_2)\f{y^5}{10}\left(1-\f{K_3^2(y)}{K_2^2(y)}\right).
\end{equation}
The retarded Green's function is given by 
\begin{equation}
G^i (\omega, T) = - \frac{1}{e^2} \lim_{r\rightarrow\infty} \left.   \frac{r^5 \phi^{i\prime}}{L^5 \phi^i} 
\right|_{\phi^j_\infty=0, j\neq i} + G_{\rm{counter}}( \omega, T ) , 
\end{equation}
with
\begin{equation}
\frac{1}{e^2} = \frac{N^3 }{ 6\pi^3 L^5 }. 
\end{equation}
Substituting the expressions for fluctuations $\phi^1$ in the $\omega\rightarrow i\infty$ limit
we obtain
\begin{equation}
G^1(\omega, T)|_{\omega\rightarrow i\infty}  =
 \f{1}{e^2} \lim_{y\rightarrow 0}
 \frac{r_+ ^5 }{L^5 y^3 } \left( \lambda^4  a_0^{(1)}(y)  + a_1^{(1)}(y)  \right) +
G_{\rm{counter}}^1 (\omega, T). 
 \end{equation}
As discussed for the D3-brane and the M2-brane we define the regularized Green's function 
as 
\begin{eqnarray}
 \delta G^i(\omega, T) &=& 
G^1(\omega, T) - G^1(\omega, 0) - \frac{r_+^4 }{e^2 L^5}\f{4}{5}(-3k_1+2k_2).
\end{eqnarray}
The constant term which is subtracted is obtained by examining the limit of 
 the first order correction $a_1^{(1)}$ given in (\ref{1stcorm5}). 
Following the same arguments developed for the D3-brane case, 
we can apply Cauchy's theorem for the regulated  Green's function 
to derive the following sum rule
\begin{equation}\label{sumrm51}
  \int_{-\infty}^\infty  \frac{d\omega}{\pi\omega}
\left(\rho^1(\omega) - \rho^1_{T=0} (\omega) \right) =
\lim_{\omega\rightarrow 0} \omega {\rm Im}\, \sigma^{1} (\omega)
- \frac{r_+^4 }{e^2 L^5}\f{4}{5}(-3k_1+2k_2), 
\end{equation}
where again the spectral density $\rho^i= {\rm Im }\, G^i$ and 
$\sigma^i$ is the corresponding conductivity. Carrying out the same procedure 
for the second component of the gauge field we obtain the sum rule 
\begin{equation}\label{sumrm52}
 \int_{-\infty}^\infty  \frac{d\omega}{\pi\omega}
\left(\rho^2(\omega) - \rho^2_{T=0} (\omega) \right) =
\lim_{\omega\rightarrow 0} \omega {\rm Im}\, \sigma^{2} (\omega)
- \frac{r_+^4 }{e^2 L^5}\f{4}{5}(2k_1-3k_2). 
\end{equation}

To understand the terms due to the high frequency limit of the Green's function 
we examine the OPE of 2 R-currents in a 6 dimensional CFT. 
This is given by 
 \begin{equation}
\label{opejj5}
 J_\mu^i(x)J_\nu^j(0) \sim \frac{ {\cal C}  \delta_{ij} I_{\mu\nu} (x)  }{x^{10}} 
+{\cal{A}}_{\mu\nu} C_{ij}^{\;\;\hat k} {\cal{O}}_{\hat k} (0) +
{\cal{B}}^{ij; \rho}_{\mu\nu; k} J^k_\rho(0) + \cdots. 
\end{equation}
In this case, the R-currents have dimension 5,  and by a simple scaling analysis
it is easy to see that the finite terms at high frequency arise from 
dimension $4$ operators in the OPE. 
For operators of dimension $4$ the regularized tensor ${\cal A}$ is given by 
\begin{equation}
 {\cal A}_{\mu\nu} (s) = - 2 
\left( \frac{1}{4} \partial_\mu\partial_\nu \frac{1}{s^4} + \pi^3 \delta_{\mu\nu}(s) \right). 
\end{equation}
Taking the Fourier transform of the OPE and going over to Minkowski space we 
obtain the following relation for the high frequency behaviour of the retarded 
Green's function
\begin{equation} \label{him5}
\lim_{\omega\rightarrow\infty} ( G^i(\omega, T) - G^i(\omega, 0)) =
2\pi^3 C_{ii}^{\; \hat k} \langle {\cal O}_{\hat k} (0) \rangle_T, 
\end{equation}
where ${\cal O}_{\hat k}$ are the operators of dimension 4 in the theory.

For the M5-branes, the operators of dimension 4 correspond to the two scalars 
$\vartheta^{\hat i}$ in the theory.  The expectation values of these scalars in the 
background (\ref{scalm5})  are given by 
\begin{eqnarray}
\langle{\cal O}_1\rangle =\frac{N^3r_+^4}{6 \pi^2 L^8}\frac{2}{\sqrt{2}} ( k_1-k_2),\nn\\
\langle{\cal O}_2\rangle =\frac{N^3r_+^4}{6 \pi^2 L^8}\frac{2}{\sqrt{10}}( k_1+k_2) .
\end{eqnarray}
Rewriting the high frequency contribution in the RHS of the sum rules given in (\ref{sumrm51}) and 
(\ref{sumrm51}) we obtain
\begin{eqnarray}
\label{srm5og}
\delta G^1(0 )= 
\lim_{\omega\rightarrow 0} \omega {\rm Im}\, \sigma^{1} (\omega)
+ \f{\sqrt{2}}{L^2}\langle{\cal O}_1\rangle +  \f{\sqrt{2}}{L^2\sqrt{5}}\langle{\cal O}_2\rangle,\nn\\
\delta G_J^2(0, T)= 
\lim_{\omega\rightarrow 0} \omega {\rm Im}\, \sigma^{2} (\omega)
- \f{\sqrt{2}}{L^2}\langle{\cal O}_1\rangle +  \f{\sqrt{2}}{L^2\sqrt{5}}\langle{\cal O}_2\rangle. 
\end{eqnarray}
We now extract the structure constants by comparing (\ref{him5}) with the high frequency behaviour
of the Green's function given in (\ref{srm5og}). 
This results in the following values for the structure constants. 
\begin{eqnarray}
\label{zsm5}
C_{11}^{\;\hat 1}  &=&-\f{\sqrt{2}}{2\pi^3L^2},\qquad 
C_{11}^{\;\hat 2}  = - \f{1}{2\pi^3L^2}\sqrt{\f{2}{5}},\nonumber\\
C_{22}^{\; \hat 1}  &=& \f{\sqrt{2}}{2\pi^3L^2},\qquad 
C_{22}^{\; \hat 2}   = - \f{1}{2\pi^3L^2}\sqrt{\f{2}{5}}.
\end{eqnarray}

Let us now compare the structure constants obtained from the sum rule in (\ref{zsm5}) with
that using the conventional AdS/CFT prescription. 
As discussed for the case of the D3-brane and the M2-brane these are given by 
\begin{equation}\label{wittm5}
C_{ii}^{\hat j} = -a_{\hat j}^{i} \frac{{\cal K}_6 (4) }{ {\cal G}_6 (4)}
\end{equation}
where the  $a_{\hat j}^{i}$ refer to the $\hat j$th component of the 
vector $ \vec a^i$ given in (\ref{scaldefm5}). 
The normalization constants in (\ref{wittm5}) are read out from 
(\ref{struc}) and (\ref{twopt}).  
Substituting these constants, we again obtain the structure constants
which were obtained from the sum rule that are given in (\ref{zsm5}).

\section{Conclusions}

We have derived the R-current spectral sum rules holographically 
for the theories dual to that of the D3-brane, M2-brane and M5-brane 
at finite chemical potential. The sum rule as we have seen is a consequence of the 
analytic behaviour of the corresponding retarded Green's function. 
It contains information about 
 both long distance hydrodynamical  property as well as the short distance 
property of the theory. 
We examine the term which occur due to the short distance effects and obtained
the relevant structure constants. As a consistency check, this was then compared to that obtained
using the conventional Witten diagrams and shown to agree. 

Thus we see sum rules provide crucial information of the theory and provide 
important constraints on the spectral densities. It will be useful to examine the 
putative holographic duals of QCD in the literature and obtain various sum rules. 
This may provide tight constraints on the validity of the holographic models  when compared 
with expectations of real  QCD sum rules. 

Anomalies are other phenomena in any quantum field theory
 which are quite robust and present both at short distance and 
long distances in the theory. 
 It will be interesting to 
determine spectral densities sensitive to the $U(1)^3$ anomaly present in
 ${\cal N}=4$ Yang-Mills hydrodynamics \cite{Son:2009tf}  and obtain sum rules 
which capture this anomaly.

\acknowledgments
We thank Sachin Jain and R. Loganayagam for useful discussions. 
The work of J.R.D is partially supported by the Ramanujan fellowship DST-SR/S2/RJN-59/2009.

\appendix
\section{ 2pt and 3pt  functions from Witten diagrams}

\vspace{.5cm}
\noindent
{ \bf The $\langle J J {\cal O} \rangle$ correlator}
\vspace{.5cm}

In this section we evaluate the three point function of two currents
 with a scalar 
operator of dimension $\Delta$ for a conformal field theory in $d$ dimensions
using the AdS/CFT correspondence.  
\begin{equation}
\label{interact}
 \frac{1}{ 4 e^2 } \int d^{d+1} w \sqrt {g} g^{\mu\rho} g^{\nu\sigma} 
\vartheta \partial_{ [ \mu } A_\nu \partial_{ \rho ] } A_\sigma  . 
\end{equation}
Here $\vartheta$ is the scalar dual to the operator of dimension $\Delta$ and 
the gauge field $A_\mu$ is dual to a current. The current in the dual theory 
has the dimension $d-1$.  
The metric for $AdS_{d+1}$ is given by 
\begin{equation}
 ds^2 = \frac{L^2}{z_0^2} ( d\vec x^2 + dz_0^2 ) .
\end{equation}
We have chosen the radius of $AdS_{d+1}$ as $L$ in agreement with the asymptotic
form of the metrics given for the charged D3-brane, M2-brane and the M5-brane
which are given in (\ref{chgmet}), (\ref{ch4met}) and (\ref{ads7}). 
The bulk to boundary Green's function for the scalar is given by \cite{Freedman:1998tz}
\begin{equation}
\label{bbgre}
 K_{\Delta} ( z_0, \vec z, \vec x) = \frac{ \Gamma(\Delta)}{\pi^{\frac{d}{2} } 
\Gamma ( \Delta - \frac{d}{2}  ) }\left( \frac {z_0}{ z_0^2 + ( \vec z - \vec x)^2 }  \right)^{\Delta}. 
\end{equation}
Note that though here we are following the normalization in
which conflicts with a Ward identity \cite{Freedman:1998tz} we rescale the final results
by using the normalization given in \cite{Klebanov:1999tb}. 
The bulk to boundary Green's function for the vectors is given by \cite{Freedman:1998tz}
\begin{equation}
 G_{\mu i}( z, \vec x) = C^d \left( \frac{ z_0}{ ( z- \vec x)^2 } \right)^{d-2} 
\partial_\mu \left( \frac{ ( z - \vec x )_i }{( z - \vec x) ^2}  \right) , 
\end{equation}
where 
\begin{equation}
 C_ d = \frac{\Gamma(d) }{2 \pi^{\frac{d}{2}} \Gamma( \frac{d}{2}) }. 
\end{equation}
Now substituting these in the interaction given in (\ref{interact}) and evaluating the 
integral using the methods of \cite{Freedman:1998tz} we obtain the following expression for the 
there point function
\begin{eqnarray}\label{struc1}
\langle J_{\mu} (x) \rangle J_{\nu} (y)  {\cal O}( z) \rangle =
-{\cal K}^d( \Delta) 
 \left( \frac{- ( \Delta - 2 (d-1) ) I_{\mu\nu} (x-y) - \Delta I_{\mu \rho} ( x-z)  I_{\rho}  ( z-y) }
{  ( x-y)^{ 2( d-1) -\Delta} ( x-z)^\Delta  ( y-z)^{\Delta} }\right).   \nonumber \\
\end{eqnarray}
The normalization constant 
 ${\cal K}$  in (\ref{struc1}) is given by 
\begin{eqnarray}
\label{struc}
 {\cal K}_{d} (\Delta)  &=& \frac{L^{d-3}}{e^2} \frac{( d-2)^2 }{16  \pi^{\frac{ d}{2} } }
\left( \frac{ \Gamma( \frac{\Delta + d}{2} -1)  \Gamma ( d-1 - \frac{\Delta}{2} ) 
\Gamma ( \frac{\Delta}{2} +1)   \Gamma ( \frac{\Delta}{2} ) }
{\Gamma (\Delta)  ( \Gamma(\frac{d}{2} )) ^2 }\right) \nonumber \\
& &\qquad \times 
\left(  \frac{ \Gamma(\Delta)}{ \pi^{ \frac{d}{2} }  \Gamma( \Delta - \frac{d}{2} +1 ) } \right) .  
\end{eqnarray}
Note that  we have  finally divided by $\Delta -d/2$ following  the normalizations of 
\cite{Klebanov:1999tb}. Note that the expression in (\ref{struc}) is valid all the way
down to the unitary bound $ d/2 -1$ and thus includes scalars which saturate the 
Breitenlohner-Freedman bound.

\vspace{.5cm}
\noindent
{\bf The $\langle {\cal O} {\cal O} \rangle $ correlator}
\vspace{.5cm}

The two point function of an scalar operator ${\cal O}$ of conformal dimension 
$\Delta$ in a $d$ dimensional theory is obtained by considering the following
Lagrangian in $AdS_{d+1}$. 
\begin{equation}
S = \frac{1}{2 e^2} \int d^{d+1} z \sqrt{g} \left(
 g^{\mu\nu} \partial_\mu\vartheta\partial_\nu \vartheta + m^2 \vartheta^2 
\right). 
\end{equation}
Substituting the bulk to boundary Green's function for the scalar given in (\ref{bbgre}) and 
evaluating the two point function using the standard AdS/CFT rules 
and following \cite{Freedman:1998tz} we obtain
\begin{equation}
\langle {\cal O }(x) {\cal O}(y) \rangle =  {\cal G}_d(\Delta)  \frac{1}{( x- y)^{2 \Delta} }, 
\end{equation}
where the normalization ${\cal G}$ is given by 
\begin{equation} \label{twopt}
{\cal G}_d( \Delta)  = \frac{ L^{ d-1} }{e^2}  \frac{2\Gamma(\Delta) }{\pi^{\frac{d}{2}} 
\Gamma( \Delta - \frac{d}{2} +1) }. 
\end{equation}
Note that here  we have divided by $(\Delta - d/2)^2$ following the normalizations 
of \cite{Klebanov:1999tb}. The expression in (\ref{twopt}) agrees with that given in 
equation (2.21) of \cite{Klebanov:1999tb}. Here again the expression in (\ref{twopt})
is valid all the way down to the unitary bound. 

\section{The M2 and M5-brane solution}

\vspace{.5cm} \noindent
{\bf M2-brane solution}
\vspace{.5cm}

The metric and the gauge field for the R-charged M2-brane with all the four charged turned on 
 is given by 
\begin{eqnarray}
\label{ch4met}
& & ds_4^2
  = \frac{16(\pi T_0 L)^2}{9u^2}{\cal H}^{1/2}
    \left( - \frac{f}{{\cal H}} dt^2 + dx^2 + dz^2 \right)
  + \frac{L^2}{f u^2}{\cal H}^{1/2}~du^2~, \\ \nonumber
& & A_{t}^{i}  = \frac{4}{3} \pi T_0 L\sqrt{ k_i\prod\limits_{i=1}^4
 (1+k_i)}~\frac{u}{H_i}~,~~~~~~
u=\f{r_+}{r},\qquad H_i = 1 + k_i u~,  \\ \nonumber
& &{\cal H}=\prod\limits_{i=1}^4 H_i,~~~~
f = {\cal H}-\prod\limits_{i=1}^4
 (1+k_i)u^{3} , \quad  T_0=\f{3r_+}{4\pi L^2 }.
\end{eqnarray}
The scalars are given by 
\begin{equation}
\label{valscalm2}
X^i = {{\cal H}^{1/4}\over H_i(u)}. 
\end{equation}
The four scalars are not independent and are  constrained by $X_1X_2X_3X_4=1$.  
The charged M2-brane is the solution of the equation of motion of the following
action 
\begin{eqnarray}
S &=& \frac{N^{3/2}\sqrt{2}}{24\pi L^2} \int d^4 x\sqrt{-g} {\cal L}, \\ \nonumber
{\cal L} &=& R - \frac{1}{2}( \partial \vec\vartheta)^2 + V(\vartheta)  - 
\frac{1}{4} \sum_{i=1}^4 e^{\vec a^i \cdot \vec \vartheta} (F^i)^2.
\end{eqnarray}
where the fields four fields $X_i$ are related to the three independent fields 
$\vec \phi_i = ( \theta_1, \theta_2, \theta_3) $  by 
\begin{eqnarray}
\label{indpsc4}
& &  X_i = \exp( - \frac{1}{2} \vec a^i \vec  \vartheta ) , \\ \nonumber
& & \vec a^1 = ( 1, 1, 1), \quad 
\vec a^2 = ( 1, -1, -1) , \quad  \vec a^3 = ( -1, 1, -1) , \quad
\vec a^4 = ( -1, -1, 1) .
\end{eqnarray}
The scalar potential is given by 
\bea
\label{sc4pot}
V(\vartheta)=\f{2}{L^2}(\cosh \vartheta_1 + \cosh \vartheta_2 + \cosh \vartheta_3).
\eea
The   chemical potentials are given by 
\begin{eqnarray}
\label{chem4}
\mu_i &=& \frac{4 \pi\, T_0}{3}L\, \frac{1}{1+k_i}\sqrt{ \, k_i \prod\limits_{i=1}^4(1 + k_i) }.\nn
\end{eqnarray}

\vspace{.5cm}\noindent
{\bf M5-brane}
\vspace{.5cm}

The metric and the gauge field for the R-charged M5-brane with all the charges turned 
on  is given by \cite{Duff:1999rk}
\begin{eqnarray}
\label{ads7}
&&ds_7^2
  = \frac{4(\pi T_0 L)^2}{9u}{\cal H}^{1/5}
    \left( - \frac{f}{{\cal H}} dt^2 + dx_1^2 + \cdots + dx_4^2 + dz^2 \right)
  + \frac{L^2}{4 f u^2}{\cal H}^{1/5}~du^2~,  \nonumber \\ 
&&A_{t}  = \frac{2}{3} \pi T_0 \sqrt{2 k_i\prod\limits_{i=1}^2(1+k_i)}~\frac{u^2}{H_i}~,~~~~~H_i = 1 + k_i u^2~,\nn\\
&&T_0=\f{3 r_+}{2 \pi L^2},\qquad  u=\frac{r_+^2}{r^2}
\\ \nonumber
&&H_i = 1 + k_i u^2~, ~~~~~~~~
{\cal H}=\prod\limits_{i=1}^2 H_i, \quad
f = {\cal H}-\prod\limits_{i=1}^2(1+k_i)u^{3}~. 
\end{eqnarray}
The background solution for the two scalars are given by 
\begin{equation}
\label{scalm5}
X^i = {{\cal H}^{2/5}\over H_i(u)}. 
\end{equation}
The chemical potentials for the solution in (\ref{ads7}) are given by 
\begin{equation}\label{chemm5}
\mu_i = \frac{2}{3} \pi T_0 \sqrt{2 k_i\prod\limits_{i=1}^2(1+k_i)}~\frac{1}{1+ k_i}. 
\end{equation}
The charged M5-brane background is a solution of the equation of motion of the 
following action 
\begin{eqnarray}
S &=& \frac{N^3}{6 \pi^3 L^5} \int d^7 x\sqrt{-g} {\cal L}, \\ \nonumber
{\cal L} &=& R - \frac{1}{2}( \partial \vec\vartheta)^2 + V(\phi)  - 
\frac{1}{4} \sum_{i=1}^2 e^{\vec a^i \cdot \vec \vartheta} (F^i)^2. 
\end{eqnarray}
Where the two  scalar fields $X_i's$ are related to $\vec{\vartheta}=(\vartheta_1,\vartheta_2)$ b;y 
\bea \label{scaldefm5}
&& X^i=e^{-\f{1}{2}\vec{a}^i\cdot \vec \vartheta },\nonumber\\
&& \vec{a^1} =(\sqrt{2},\sqrt{\f{2}{5}}),\qquad 
\vec{a^2}=(-\sqrt{2},\sqrt{\f{2}{5}}). 
\eea
Note that the two scalars $X_i$ are independent here, unlike the situation in 
the case of the D3-branes and M2-branes.
The  scalar potential  $V$ is given by 
\be
\label{m5pot}
V= \frac{4}{L^2} \left( -4X_1 X_2-2X_1^{-1}X_2^{-2}-2X_1^{-2}X_2^{-1}+\f{1}{2}(X_1 X_2)^{-4} \right).
\ee

\providecommand{\href}[2]{#2}\begingroup\raggedright\endgroup

\end{document}